\documentclass[11pt,a4paper]{article}
\usepackage{graphicx,amsfonts,amsmath,amssymb,bm} 
\usepackage{a4wide,authblk,xcolor,multirow}
\usepackage[linktocpage=true,colorlinks=true,linkcolor=brown, pdfborder={0 0 0}]{hyperref}
\usepackage{booktabs}   
\usepackage{tabularx}   
\usepackage{makecell}   
\usepackage{array}      
\newcolumntype{L}{>{\raggedright\arraybackslash}X} 
\usepackage{multirow}

\definecolor{augustine}{rgb}{1,0,1}

\begin{document}

\title{\bfseries Quantum Advantage: a Tensor Network Perspective}

\author[1]{Augustine Kshetrimayum\thanks{\href{mailto:augustine.kshetrimayum@multiversecomputing.com}{augustine.kshetrimayum@multiversecomputing.com}}}
\author[1,2]{Saeed S.~Jahromi}
\author[3]{Sukhbinder Singh}
\author[1,2,4]{Rom\'{a}n Or\'{u}s}

\affil[1]{Multiverse Computing, Parque Cientifico y Tecnol\'{o}gico de Gipuzkua, Paseo de Miram\'{o}n, 170 3$^{\,\circ}$ Planta, 20014 Donostia / San Sebasti\'{a}n, Spain}
\affil[2]{Donostia International Physics Center, Paseo Manuel de Lardizabal 4, E-20018 San Sebasti\'an, Spain}
\affil[3]{Multiverse Computing, Centre for Social Innovation, 192 Spadina Avenue Suite 509, Toronto, ON M5T 2C2, Canada}
\affil[4]{Ikerbasque Foundation for Science, Maria Diaz de Haro 3, E-48013 Bilbao, Spain}

\date{\today}

\maketitle
\begin{abstract}
    We review the recent quantum advantage experiments by IBM, D-Wave, and Google, focusing on cases where efficient classical simulations of the experiment were demonstrated or attempted using tensor network methods. We assess the strengths and limitations of these tensor network–based approaches and examine how the interplay between classical simulation and quantum hardware has advanced both fields. Our goal is to clarify what these results imply for the next generation of quantum advantage experiments. We identify regimes and system features that remain challenging for current tensor network approaches, and we outline directions where improved classical methods could further raise the standard for claiming quantum advantage. By analyzing this evolving competition, we aim to provide a clear view of where genuine, scalable quantum advantage is most likely to emerge.
\end{abstract}
\newpage 
\enlargethispage{1cm}
\tableofcontents
\clearpage

\section{Introduction}
\label{sec:intro}
\emph{Quantum advantage} or \emph{quantum supremacy} is a term used to denote the situation where quantum processors can outperform classical computers for some specific problems~\cite{QuantumsupremacyPreskill}. The performance could be in terms of accuracy of a particular solution, efficiency in obtaining those solutions, energy consumption of the devices while trying to obtain those solutions, etc. Similar definitions include the ability of a quantum computer to perform tasks that no classical computers can, in any reasonable amount of time. \emph{Quantum utility} is another term used to describe the practical usefulness of a quantum computer~\cite{QuantumUtilityIBM,QuantumutilityHerrmann2023}. While the emphasis shifts from strictly outperforming classical devices to demonstrating meaningful practical value, the concept remains closely related to quantum advantage or quantum supremacy. In practice, a quantum processor cannot be regarded as genuinely useful if a classical computer can accomplish the same task with comparable or fewer resources, superior performance, or both.
Some have extended the sense of quantum advantage to also include hybrid computational setups --- using quantum hardware along with classical supercomputers (consisting of CPUs,  GPUs, etc.) to outperform purely classical computation in some way or the other~\cite{quantumCentricIBM}.

Since the realization of a universal fault-tolerant quantum computer is still likely years away, increasing attention has shifted toward demonstrating quantum advantage with current noisy and resource-limited devices in the so-called \emph{Noisy Intermediate Scale Quantum} (NISQ) era~\cite{nisqPreskill}. Despite current technical limitations and challenges, most quantum computing platforms have placed the demonstration of some form of quantum advantage prominently on their near-term roadmaps, reflecting growing confidence in the maturity of the technology. For instance, IBM has outlined plans to demonstrate quantum advantage for scientific applications by 2026~\cite{quantumroadmapIBM}. Other platforms, including D-Wave Systems~\cite{dwavesupremacy2025} and Google~\cite{Google_quantum_echo2025}, have recently reported quantum advantage claims, which are currently under scrutiny by the broader research community~\cite{QadvantagereviewEisert2025, Tindalldisordyn2d3d, qadvantagetracker, HangleiterQadvantage2026}. 

Some of the world's major quantum computers are listed in Table \ref{tab:qc_hardware}.  There have been three main types of quantum advantage experiments showcased on some of these platforms:(1) Gaussian Boson Sampling (GBS), (2) Random Circuit Sampling (RCS), (3) Simulation of the non-equilibrium dynamics of a quantum many-body system. We focus our discussions on quantum simulation experiments by IBM~\cite{IBMkickedisingNat2023} and D-Wave~\cite{dwavesupremacy2025} and RCS experiments by Google~\cite{GooglesupremacyRCSNat2018,Google_quantum_echo2025}. We do not discuss GBS experiments in this work~\cite{GBS_Jiuzhang2020, GBSXanadu2022} mainly because of its verification hardness~\cite{GBSPRX_XEcertficiation2024, carolan_experimental_2014} and lack of practical applications beyond sampling~\cite{BSapplicationsPRA2016} (in contrast, RCS has been recently applied to probe quantum chaos \cite{Google_quantum_echo2025}).

\clearpage
\begin{table}[htbp]
\centering
\small
\setlength{\extrarowheight}{2pt}

\begin{tabular}{|
    p{2.1cm}  
  | p{2.0cm}  
  | p{2.0cm}  
  | p{2.0cm}  
  | p{3.0cm}  
|}
\hline
\textbf{Organization} & \textbf{System} & \textbf{Qubits} & \textbf{Hardware} & \textbf{Connectivity} \\
\hline

\multirow{2}{*}{Google} 
 & Sycamore & 53--54 &
   \makecell[l]{Transmons} &
   \makecell[l]{2D square lattice}
 \\ \cline{2-5}
 & Willow & 105 &
   \makecell[l]{Transmons} &
   \makecell[l]{2D square lattice}
 \\ \hline

\multirow{2}{*}{IBM} 
 & Eagle & 127 &
   \makecell[l]{Transmons} &
   \makecell[l]{Heavy-hex lattice}
 \\ \cline{2-5}
 & Osprey & 433 &
   \makecell[l]{Transmons} &
   \makecell[l]{Heavy-hex lattice}
\\ \cline{2-5}
 & Condor & 1121 &
   \makecell[l]{Transmons} &
   \makecell[l]{Heavy-hex lattice}
 \\ \hline


\multirow{2}{*}{UST, China} 
 & Zuchongzhi 2.1
 & 66 &
 \makecell[l]{Transmons} &
 \makecell[l]{$11 \times 6$ grid}
 \\ \cline{2-5}

 & Zuchongzhi-3.0
 & 105 &
 \makecell[l]{Transmons} &
 \makecell[l]{$15 \times 7$ grid}
 \\ \hline

\multirow{2}{*}{UST, China} 
 & Jiuzhang 1.0
 & 76 photons / 100 modes &
 \makecell[l]{Photonic \\(GBS)} &
 \makecell[l]{--}
 \\ \cline{2-5}

 & Jiuzhang 2.0
 & 113 photons / 144 modes &
 \makecell[l]{Photonic\\ (GBS)} &
 \makecell[l]{--}
 \\ \hline

\multirow{1}{*}{D-Wave}
 & Advantage2 &
   $>5000$ &
   \makecell[l]{Flux-qubit\\annealer} &
   \makecell[l]{Pegasus/Zephyr\\sparse graph}
 \\ \hline

\multirow{1}{*}{Xanadu}
 & Borealis & 216 modes &
   \makecell[l]{Photonic CV} &
   \makecell[l]{Dense interferometer\\graph}
 \\ \hline

\multirow{1}{*}{Quantinuum}
 & H2-1 & 56 &
   \makecell[l]{Trapped ions} &
   \makecell[l]{All-to-all native\\connectivity}
 \\ \hline

\multirow{1}{*}{IonQ}
 & Aria/Forte & 31--32 &
   \makecell[l]{Trapped ions} &
   \makecell[l]{All-to-all \\connectivity}
 \\ \hline

\multirow{1}{*}{Rigetti}
 & Aspen-M & 80 &
   \makecell[l]{Transmons} &
   \makecell[l]{2D Square lattice}
 \\ \hline

\multirow{1}{*}{QuEra}
 & Aquila & 256 &
   \makecell[l]{Neutral atoms\\(Rydberg)} &
   \makecell[l]{Flexible 2D\\atom array}
 \\ \hline

\multirow{1}{*}{Pasqal}
 & Fresnel &
   100--1000 &
   \makecell[l]{Neutral atoms\\(Rydberg)} &
   \makecell[l]{Flexible 2D\\arrays}
 \\ \hline

\end{tabular}
\caption{Major quantum computing platforms. Connectivity corresponds to how qubits are physical arranged on the hardware; quantum gates are applied along the links of this geometry. UST: University of Science and Technology, China; GBS: Gaussian Boson Sampling; CV: Continuous Variables.}
\label{tab:qc_hardware}
\end{table}
Despite the recent claims by different quantum computing platforms, achieving quantum advantage is not a straightforward task that can be addressed simply by improving a specific piece of quantum hardware. Any meaningful demonstration of quantum advantage must contend with strong competition—not only from other quantum platforms, but also from classical hardware and algorithms. One cannot credibly claim an advantage without benchmarking against the best available state-of-the-art classical methods~\cite{Pasqal_clas_sim2025}. However, identifying the optimal classical algorithm for a given quantum problem is often nontrivial and, in many cases, remains an open and actively researched question. As a result, claims of quantum advantage are inherently tied to the evolving landscape of classical algorithms. One of the objectives of this article is therefore to highlight this issue and to emphasize the necessity of carefully defined and continually updated classical benchmarks with a special focus on Tensor network methods~\cite{orus_TNreivew_2014, VerstraeteTNreview2008, EisertTNreview2013}. This emphasis is motivated by two main reasons. First, tensor networks have emerged as some of the most powerful and versatile tools for generally studying strongly correlated quantum many-body systems, a problem area that has also become directly relevant for benchmarking quantum computers~\cite{TNforquantumcomputing2025, noisy_circuit_Cichy2025}. Second, many recent quantum advantage experiments have in practice adopted Tensor network techniques as part of their classical benchmarking strategies.

The remainder of the paper is organized as follows. In Sec.~\ref{sec:TNs}, we briefly review the tensor network methods that have been applied in quantum advantage studies. In Sec.~\ref{sec:Qadvantageexperiments}, we review two prominent quantum advantage experiments in the context of quantum simulation, one performed by IBM~\cite{IBMkickedisingNat2023} and the other by D-Wave~\cite{dwavesupremacy2025}. This is followed by Google's Random Circuit sampling experiment~\cite{GooglesupremacyRCSNat2018} and their subsequent related work on quantum echoes~\cite{Google_quantum_echo2025}. We describe their experimental setups and the benchmarking results obtained in comparison with state-of-the-art tensor network approaches. We then discuss several follow-up works based on tensor network simulations that challenge or refine some of these claims. In Sec.~\ref{sec:lims_directions}, we provide a broader discussion of these results in the context of future quantum advantage claims, highlighting regimes where tensor network methods face intrinsic limitations and which may therefore offer promising opportunities for demonstrating quantum advantage with existing hardware. We also comment on future directions, including how tensor network methods are expected to continue improving through the use of GPUs and modern machine learning techniques. Finally, we conclude with a discussion on how the competition between tensor networks and quantum computers will continue to shape the field.

\begin{figure}
	\begin{center}
		\includegraphics[width=0.9\textwidth]{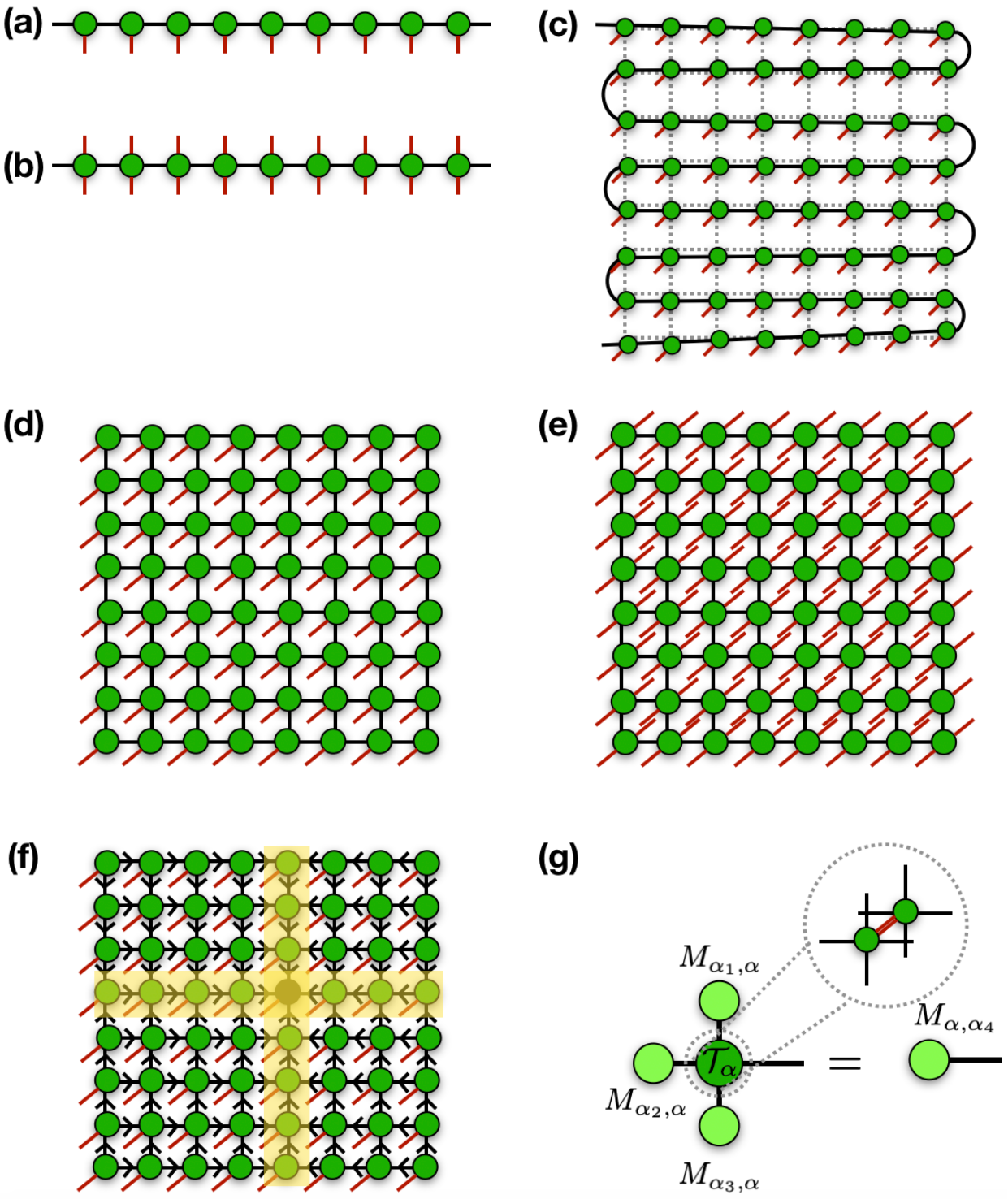}
		\caption{ (a) Matrix Product States (MPS), (b) Matrix Product Operators (MPOs), (c) MPS wrapped in a snake pattern to simulate a 2d lattice, (d) Projected Entangled Pair States (PEPS), (e) Projected Entangled Pair Operators (PEPO), (f) Example of an orthogonality centre in a 2d isometric Tensor network, (g) Self consistent equation for defining the message tensors in Belief Propagation (BP) algorithm.
		}
	\end{center}
	\label{fig:TNdiagrams}
\end{figure}

\section{Tensor network methods for quantum advantage experiments}
\label{sec:TNs}
Tensor network techniques~\cite{orus_TNreivew_2014, VerstraeteTNreview2008, EisertTNreview2013, tebdVidal2004, dmrgWhitePRB1993}, originally developed for the study of strongly correlated problems in condensed matter physics, have continued to exceed expectations in terms of both their capabilities and range of applications. One of their most prominent recent applications is in quantum computing~\cite{TNforquantumcomputing2025}. Today, a broad and rapidly growing family of tensor network algorithms exists, addressing problems that span quantum simulation, quantum information, and even machine learning.
In this section, we review only the most popular tensor network methods that have been employed to analyze and benchmark claims of quantum advantage. At the same time, the challenge posed by quantum advantage experiments is steadily strengthening tensor network methods and is driving the development of new tensor network techniques.

\subsection{Matrix Product States}
Matrix product states (MPS) are the most widely used tensor network states and arguably the basis of the most developed suite of algorithms for simulating 1D and quasi 1D quantum lattice systems. We refer the reader to several excellent review articles on MPS for further details \cite{schollwock_MPSreview_2011,orus_TNreivew_2014}. An MPS consists of a network of tensors arranged in 1D, as illustrated in Fig.~\ref{fig:TNdiagrams}~(a). This ansatz is particularly well suited for approximating low-lying eigenstates of 1D local Hamiltonians that obey the so-called area law of entanglement \cite{RMP_arealaw_Eisert2010}. Each tensor in an MPS carries one physical index, which corresponds to the Hilbert space of a lattice site, and either one or two bond indices that connect it to neighboring tensors. This tensor network forms the basis of several different simulation algorithms. For example, the density matrix renormalization group (DMRG) is a powerful method for approximating the ground state and other low-lying eigenstates of 1D Hamiltonians \cite{dmrgWhitePRL1992,dmrgWhitePRB1993} as MPS. DMRG is based on a variational principle that minimizes the energy within the MPS manifold. Its time-dependent extension, commonly referred to as t-DMRG, can be used to simulate real-time dynamics \cite{daley_tDMRG_2004}. The time-evolving block decimation (TEBD) technique is another widely used MPS-based method for approximating the ground state of 1D local Hamiltonians \cite{tebdVidal2004, tebdPRLVidal2003}. The time-evolution operator (exponential of the Hamiltonian) is Trotterized into a sum of one- and two-body gates, which are then repeatedly applied to an initial MPS, driving the initial state toward the ground state in imaginary time. The same algorithm can also be used to simulate dynamics by performing real-time evolution (Trotterizing the real-time evolution operator) instead of imaginary-time evolution. Another powerful MPS-based approach is the time-dependent variational principle (TDVP) \cite{tdvp_verstraetePRL2011,tdvp_verstraete2016}. This method projects the action of the Hamiltonian onto the tangent space of the MPS manifold, thereby providing an effective way to solve the time-dependent Schrödinger equation within the variational ansatz. TDVP can be employed for both real-time and imaginary-time evolution. Although MPS are intrinsically 1D tensor network states, they have also been successfully applied to the study of 2D systems \cite{stoudenmire_2d_dmrg2012}. Owing to their flexibility, maturity, and computational efficiency, MPS-based methods remain the most commonly used classical techniques in quantum-advantage experiments, including those that we discuss in this paper \cite{IBMkickedisingNat2023, dwavesupremacy2025, GooglesupremacyRCSNat2018, Google_quantum_echo2025}.

Matrix product operators (MPOs) are the operator counterparts of MPS and share many of their properties. They can be used to represent density matrices as well as many-body operators, including Hamiltonians. MPO representation of local Hamiltonians (also employed in DMRG) often help mitigate the simulation cost of dynamics. In the context of quantum advantage experiments MPOs have been employed, for example, to study time evolution in the Heisenberg picture for benchmarking IBM’s kicked Ising experiment \cite{IBMkickedisingNat2023}. A schematic diagram of MPS and MPO tensor networks is shown in Fig.~\ref{fig:TNdiagrams}(a, b). In Fig.~\ref{fig:TNdiagrams}(c), we illustrate how MPS or MPOs can be used to study 2D systems by mapping them onto a 1D structure using a snake-like arrangement of 2D lattice sites.

\subsection{Projected Entangled Pair States}
Projected entangled pair states (PEPS) are tensor network states defined directly on a higher-dimensional lattice \cite{VerstraeteTNreview2008, VerstraetePEPS2004, orus_TNreivew_2014}. Unlike MPS-based approaches for 2D systems, PEPS do not require mapping a short-range Hamiltonian onto an effective long-range one. In this sense, PEPS preserve the local interaction structure of the Hamiltonian and are therefore a more natural ansatz for representing 2D (and higher-dimensional) quantum states. PEPS are most commonly defined on a square lattice, where each local tensor has four bond indices and one physical index --- containing a total of $D^4p$ parameters --- where $D$ is the bond dimension that quantifies the amount of entanglement between neighboring sites and $p$ the physical dimension of the lattice Hilbert space. Analogous to MPOs, projected entangled pair operators (PEPOs) are the operator counterparts of PEPS and can be used to represent Hamiltonians or density operators acting on higher-dimensional lattices~\cite{Kshetrimayum_NatComm_2017, WeimerRMP2021}. PEPOs have also been employed to study quantum dynamics and to benchmark IBM’s kicked Ising experiment \cite{IBMkickedisingNat2023}. A schematic diagram of PEPS and PEPOs is shown in Fig.~\ref{fig:TNdiagrams}(d, e).

\subsubsection{Standard PEPS}
By standard PEPS, we refer to PEPS in their simplest and most unconstrained form (e.g., without any gauge fixing or imposed isometric conditions). Even in this basic formulation, PEPS can already be very powerful in many situations. However, the computational cost is significantly higher than in the 1D case. On a square lattice, it can scale as high as $O(D^{13})$ with the bond dimension, compared to $O(D^3)$ for an MPS on a 1D lattice. This high cost arises mainly from computing the environment of a PEPS tensor—that is, the tensor obtained by contracting the rest of the network surrounding the given tensor. Most PEPS algorithms reduce this cost by replacing the exact environment computation with an approximate scheme. 

One particularly drastic approximation is the so-called \emph{simple update}, in which tensor environments are largely ignored and instead approximated by a product of operators, one acting on each bond index of the tensor. Remarkably, despite this extreme simplification, the method has been shown to perform reasonably well in several simulations of ground states, dynamics and open 2D systems~\cite{simpleupdateXiang2008, XiangKHAF2017, Kshetrimayum2dMBL2020, Kshetrimayum2dThermal2019, Kshetrimayum_NatComm_2017, SchmollfinTmaterials2024, kshetrimayum_materialsAOP2020}. This technique relies on a 2D generalization of MPS-based TEBD algorithm for updating the PEPS, while employing mean-field–like (tensor product) environments. Expectation values can be computed either within this mean-field approximation or by using more accurate, but also more costly, environment approximation schemes such as \emph{boundary MPS technique} \cite{LubaschfinPEPS2014, VerstraeteTNreview2008} or \emph{corner transfer-matrix renormalization group (CTMRG)} \cite{ctmrg_Nishino1996,ctmrgOrus2009}. 

The simple update method has been adapted to PEPS defined on arbitrary geometries (graph-PEPS) \cite{JahromigPEPS2019, PatragPEPS2024}, and has been used to benchmark quantum-advantage experiments \cite{PatragPEPS2024}, including the IBM kicked Ising experiment \cite{IBMkickedisingNat2023} and the annealing experiment by D-Wave~\cite{dwavesupremacy2025}. We will discuss these developments in more detail later. More sophisticated variants of this approach that incorporate the full environment (full update) \cite{VerstraetePEPS2004, orus_TNreivew_2014}, or parts of the environment (cluster update or neighborhood tensor update) \cite{LubaschfinPEPS2014, NTU_Dziarmaga2021}, are also available. These methods have likewise been employed to benchmark quantum-advantage experiments, such as those performed by D-Wave \cite{dwavesupremacy2025}.

\subsubsection{PEPS with Belief Propagation}
The simple update for PEPS has been formalized \cite{pepsbp} in terms of Belief Propagation (BP), a message-passing algorithm widely used for approximate inference and the computation of local marginals in probabilistic graphical models \cite{BP_Pearl1982}. In the PEPS setting, BP can be used to improve --- in some sitations --- the accuracy of the simple update by identifying more favorable gauge choices on the bond indices \cite{TindallBP2023}. The approach proceeds by determining a set of so-called message tensors acting on the PEPS bond indices by solving a system of equations for the network as illustrated in Fig.~\ref{fig:TNdiagrams}~(g). These equations are numerically solved by means of an iterative procedure (``locally passing the message tensors through the PEPS tensors''), which may or may not converge. When the procedure converges, the converged message tensors provide an approximation to the environment tensors, which can be used both during time evolution of the PEPS as part of a simple update approximation and for computing expectation values of local observables. When the BP approximation is valid, one can access very large bond dimensions both during time evolution and in the evaluation of observables. In particular, the message-passing procedure itself is highly efficient --- it involves only local contractions between messages and PEPS tensors and therefore requires relatively little computational cost. A key challenge is to characterize the regimes in which the BP approximation is reliable. Notably, BP becomes exact for PEPS defined on tree graphs. By extension, one expects the approximation to perform well for states whose correlation structure is approximately tree-like.

BP-based techniques have emerged as a powerful and increasingly popular tool within the tensor network family for benchmarking quantum-advantage experiments, including those performed by IBM and D-Wave \cite{TindallBPPRX2024, Tindalldisordyn2d3d}. Moreover, these methods are rapidly evolving, with ongoing developments aimed at overcoming the limitations of the simplest BP approximations~\cite{LoopBPEvenbly2024}.

\subsubsection{Isometric Tensor network states}
Isometric tensor network states (isoTNS) in 2D constitute a special class of PEPS introduced with the goal of enabling a more efficient computation of the environment tensors \cite{ZalatelisoTN2020}. In these PEPS, each tensor satisfies an isometric condition such that contracting a row or a column of the 2D tensor network reduces it to the so-called canonical form of an MPS, which can then be contracted efficiently. This construction can be achieved in two ways: (i) by first finding an MPS representation of the 2D wavefunction and subsequently transforming it into an isoTN, or (ii) by directly applying a 2D version of TEBD along the rows and columns of the lattice. Both procedures require moving the orthogonality center within the 2D network, a process that can be carried out using the so-called Moses move, as shown in Fig.~\ref{fig:TNdiagrams}(f). The isometric tensor network technique has been employed to benchmark IBM’s quantum-advantage experiment \cite{IBMkickedisingNat2023}.

\section{Quantum advantage experiments}
\label{sec:Qadvantageexperiments}
In this section, we briefly review the quantum advantage experiments and refer interested readers to the original publications for in-depth descriptions. We focus here on analyzing these experiments along with their tensor network simulations and related follow up works.  In what follows, we briefly review three experimental setups that reported claims of quantum advantage, together with an assessment of the tensor network simulations used to benchmark their results.

\subsection{IBM}
IBM’s quantum-computing platform is based on superconducting processors built from transmon qubits~\cite{IBMQuantum}. Over the past several years, IBM has developed a succession of increasingly large and sophisticated devices, ranging from early processors such as Eagle (127 qubits) to larger chips including Osprey (433 qubits) and Condor (1,121 qubits), as well as more recent designs such as Heron, which emphasize improved coherence and noise mitigation. In parallel, IBM has introduced modular architectures, most notably IBM Quantum System Two, which enable the integration of multiple processors into a unified platform. All of these systems operate within complex cryogenic environments and are accessed through IBM’s cloud-based infrastructure, with experiments typically programmed using the open-source Qiskit framework~\cite{IBMQiskit}.
We discuss a recent experiment that was particularly relevant for quantum advantage claims along with associated Tensor network simulations below. 

\subsubsection{The kicked Ising experiment}
In 2023, the IBM team reported results on the real-time dynamics of a 2D transverse-field Ising model implemented on their Eagle 127-qubit superconducting quantum processor~\cite{IBMkickedisingNat2023}, arranged on a heavy-hexagon lattice described by the following Hamiltonian
\begin{equation}
H = - J \sum_{\langle i, j \rangle} \sigma^z_i \sigma^z_j + h \sum_i \sigma^x_i .
\end{equation}
After Trotterizing the time-evolution operator into two-body rotation gates $e^{-iH_{ZZ}\delta t}$, corresponding to the interaction term, and single-qubit rotations $e^{-iH_X \delta t}$, corresponding to the transverse-field term, the authors focused on the parameter regime $2J\delta t = \pi /2$, chosen primarily for experimental simplicity. To mitigate circuit noise, the experiment employed a zero-noise extrapolation (ZNE) scheme, selected mainly for its relatively low sampling cost. The quench dynamics were implemented on the 127-qubit Eagle processor for both Clifford circuits (when $\theta_h$ is a multiple of $\pi / 2$) and non-Clifford circuits, starting from the initial product state $|0\rangle^{\bigotimes 127}$. By restricting the circuit depth to five Trotter steps, the authors measured three observables: the global magnetization $M_z$, as well as weight-10 and weight-17 observables, as defined in their paper \cite{IBMkickedisingNat2023}. Exact classical simulations of the global magnetization and the weight-10 observables were feasible at this circuit depth due to the restricted light cone and the shallow nature of the circuit, and were found to be in good agreement with the error-mitigated experimental results. For the weight-17 observable, however, exact classical simulation was no longer possible, and the authors therefore resorted to tensor network methods which we discuss below.

\subsubsection{Tensor network simulations of the IBM experiment}
The authors in~\cite{IBMkickedisingNat2023} employed matrix product states (MPS) and isometric 2D tensor networks with bond dimensions $\chi = 1024$ and $\chi = 12$, respectively, to benchmark the full 127-qubit circuit. While these approaches reproduced the experimental results in certain parameter regimes, they exhibited noticeable accuracy issues as the value of $\theta_h$ was increased, particularly in the vicinity of the Clifford point. The authors attributed the breakdown of these tensor network simulations to the growing entanglement generated by the circuit, which exceeded the representational capacity of their tensor network ansatze.

The authors then modified the circuit by adding an additional layer of single-qubit Pauli rotations, thereby eliminating the possibility of circuit-depth reduction based on light-cone arguments and rendering exact classical simulation infeasible. They reported that even MPS simulations with bond dimension $\chi = 3072$ failed to reproduce the results of the 68-qubit Light-cone and Depth Reduced (LCDR) circuit for the weight-17 observable, while simulations based on isometric tensor networks were no longer computationally feasible. The study was further extended by increasing the circuit depth to 20 Trotter steps and measuring a weight-1 observable, $\langle \sigma^z_{62} \rangle$, for different values of $\theta_h$. In this case, MPS simulations were found to agree well with the experimental data at small values of $\theta_h$, corresponding to a weakly entangling regime. However, as $\theta_h$ was increased, increasingly significant deviations from the experimental results were observed. Importantly, the agreement with the experimental data improved systematically with increasing bond dimension $\chi$, indicating that the dominant source of error was the finite bond dimension of the MPS ansatz rather than an intrinsic algorithmic failure. Based on this scaling behavior, the authors estimated that an exact classical simulation of the circuit at this depth for $\theta_h = \pi/2$ would require a bond dimension of order $10^{13}$. Such a requirement would translate into an unrealistic memory footprint, far beyond what is achievable with current classical computing resources. The authors also reported that the runtime of the classical simulations for the modified circuit was at least twice that of the quantum device required to obtain a single data point, with further performance improvements anticipated on the quantum hardware side. Based on these observations, they concluded that their 127-qubit quantum processor exhibits a performance advantage over state-of-the-art classical simulations, even in the absence of full fault tolerance.
\begin{figure}
	\begin{center}
		\includegraphics[width=1\textwidth]{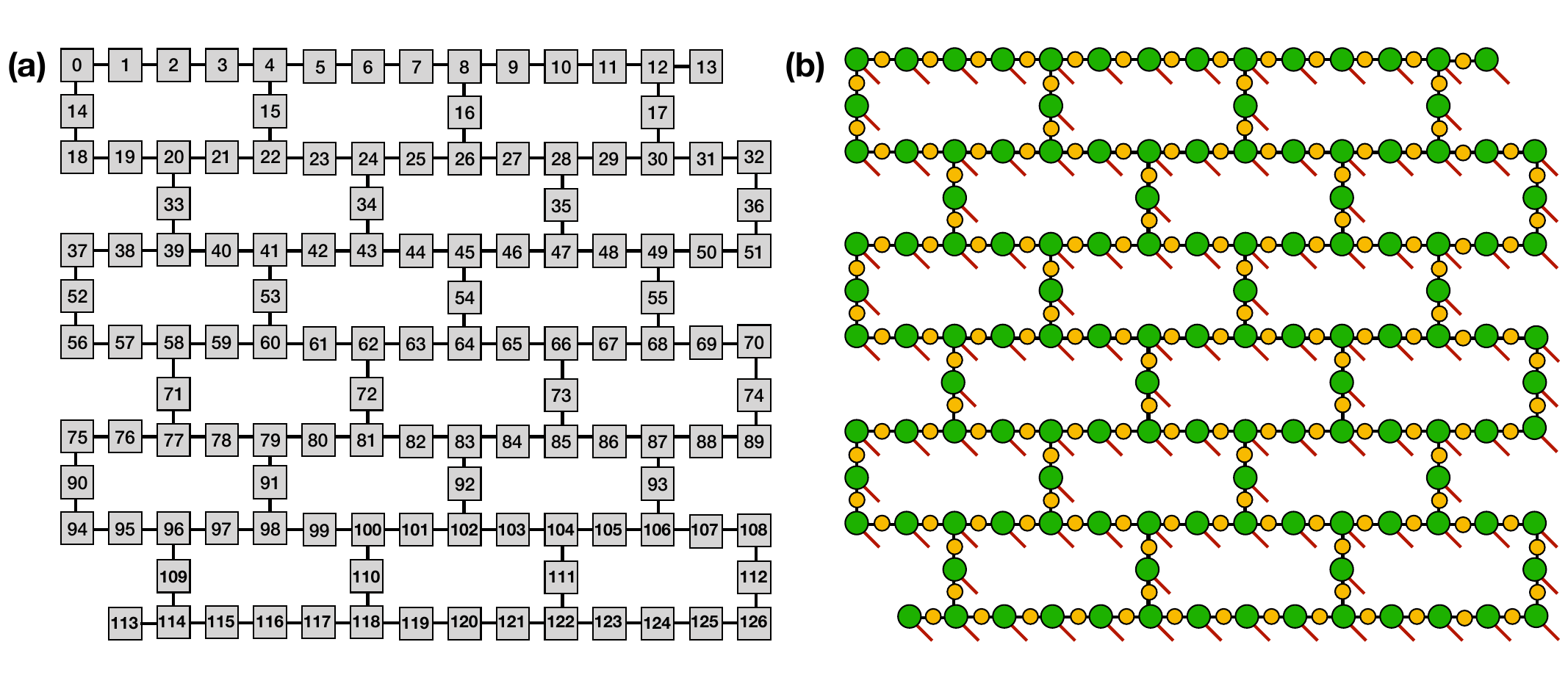}
		\caption{(a) Connectivity structure of IBM's Eagle quantum processor with 127 qubits in a heavy hexagon lattice, (b) Corresponding Tensor network structure used to simulate the processor directly mimicking the lattice structure of the quantum processor. A notable feature of this lattice --- which has been used in belief-propagation–enabled PEPS simulations --- is its locally tree-like structure: starting from a given site, one can move several steps in any direction without encountering a loop.}
        \label{fig:ibmqpu}
	\end{center}
\end{figure}

The above claim of quantum advantage was short-lived, as several subsequent works demonstrated more efficient and more accurate classical simulations of IBM’s quench experiment. In particular, the authors of \cite{TindallBPPRX2024} employed a tensor network approach with the same geometry (Fig.~\ref{fig:ibmqpu} (b)) and the same number of qubits as IBM’s quantum processor (Fig.~\ref{fig:ibmqpu} (a)), combined with the Belief Propagation algorithm, to achieve classical simulations that outperformed the quantum device in both efficiency and accuracy. By using BP for both time evolution and the computation of observables, they found that the accuracy of the method improved with increasing system size, allowing results to be obtained directly in the thermodynamic limit. To enable a direct comparison with the quantum processor, the authors simulated 127 qubits of the same model for five Trotter steps at different values of $\theta_h$ and computed single-site observables such as the magnetization $M_z$. Their results agreed with brute-force classical simulations of the corresponding LCDR circuits to an accuracy of $10^{-14}$, while requiring less than 10 seconds of runtime on a laptop. Even for higher-weight observables, such as weight-10 and weight-17 operators, the BP-based tensor network simulations achieved significantly better accuracy and runtime than both the quantum-processor results and the classical simulations reported in \cite{IBMkickedisingNat2023}. The remarkable performance of these methods was further demonstrated for the modified circuit with an additional single-qubit Pauli rotation gate (which we refer to as the $n=6$ circuit), where accuracies of up to $10^{-4}$ relative to exact results were obtained for all values of $\theta_h$ \cite{XiangPEPOIBM2023}. These results show that tensor network methods based on belief propagation decisively outperformed the quantum processor for this task, thereby refuting the original claims of quantum advantage. The authors attributed the success of their approach to the tree-like correlation structure present in the system, a point to which we will return later. For the deeper $n=20$ circuit, where exact classical solutions are not available, the authors employed finite–bond-dimension extrapolation. While they successfully reproduced the exact values at the Clifford points $\theta_h = 0$ and $\theta_h = \pi/2$, they observed discrepancies with both the hardware results and other classical methods in the intermediate regime $\pi/8 \leq \theta_h \leq 3\pi/8$.

A similar achievement in outperforming the quantum processor was reported in \cite{PatragPEPS2024} using the so-called graph-based projected entangled pair states (gPEPS), originally introduced as a generalization of iPEPS techniques to arbitrary lattice geometries in the thermodynamic limit \cite{JahromigPEPS2019}. The algorithm employed the simple update scheme for time evolution and a mean-field environment for the computation of expectation values. A key difference from the belief-propagation approach of Ref.~\cite{TindallBPPRX2024} discussed above is that the tensors are not necessarily and explicitly gauged into the favorable Vidal canonical form, however, being based on simple-updates the PEPS tends to steer towards an approximate Vidal gauge during the course of the simulation. Despite the absence of explicit gauging (which also incurs additional computation cost), the authors reported extremely high accuracy for the single-site magnetization $M_z$, reaching up to $10^{-15}$ agreement with exact simulations of the LDCR circuit for $n=5$ Trotter steps. With an average runtime of approximately 2 seconds per data point, their method outperformed not only the 127-qubit Eagle quantum processor but also the BP-based tensor network approach in both accuracy and runtime. Similar performance gains over the quantum processor were observed for higher-weight observables across different values of $\theta_h$, as well as for the $n=6$ Trotter-step circuit using bond dimensions $\chi = 32$ and $\chi = 64$. As expected, the accuracy of the tensor network results further improved with increasing PEPS bond dimension. For the deeper $n=20$ Trotter-step circuits, the authors reproduced exact results at $\theta_h = 0$ and $\theta_h = \pi/2$ using a bond dimension of $\chi = 64$. They additionally performed finite-entanglement scaling with respect to the bond dimension and observed convergence for $\theta_h \leq 3\pi/16$, beyond which noticeable deviations began to appear.

While the classical techniques discussed above are based on the Schrödinger picture of time evolution—namely, the explicit evolution of the wave function—a complementary approach was explored by evolving the operators themselves using matrix product operators (MPOs) in the Heisenberg picture \cite{ZalatelMPOIBM2023} in which operators evolve in time rather than the quantum state, according to
\begin{equation}
\langle \hat{O}(t) \rangle = \langle \psi|\hat{O}(t)|\psi \rangle ,
\end{equation}
with the time-dependent Heisenberg operator given by $\hat{O}(t) = e^{iHt}\hat{O}e^{-iHt}$. In practice, the operator $\hat{O}(t)$ is approximated by an MPO, which is then mapped to an MPS with an enlarged Hilbert space via vectorization. This MPS is subsequently wrapped around the 2D heavy-hexagon lattice using a snake-like pattern. The unitary time-evolution operator is applied by decomposing it into multiple layers in order to minimize the growth of the bond dimension, and a two-site variational matrix-product compression is employed for truncation. One notable advantage of the MPO approach is that it requires only a bond dimension $\chi = 1$ to reproduce the exact result at the Clifford point $\theta_h = \pi/2$, while still producing accurate results for other values of $\theta_h$ at circuit depth $n=5$. As a result, the MPO method performs significantly better than state-based MPS simulations, particularly in the vicinity of Clifford points. 
For circuits of depth $n=5$, the authors reported good agreement with brute-force classical simulations using LCDR (up to $10^{-4}$ accuracy) as well as with results obtained from IBM’s quantum hardware. Notably, the MPO representation required only a bond dimension $\chi = 1$ to reproduce the exact result at the Clifford point $\theta_h = \pi/2$. For the $n=6$ circuit, the MPO simulations were again found to be in good agreement with the quantum-hardware results. At a circuit depth of $n=20$, while several numerical techniques continued to agree within certain parameter regimes, discrepancies between different methods emerged for $\theta_h > \pi/8$. The Heisenberg-picture time evolution was subsequently extended to 2D tensor networks using projected entangled pair operators (PEPOs) in \cite{XiangPEPOIBM2023}. In that work, the authors further developed an extension of Clifford theory for the $n=6$ circuit, rendering it exactly solvable. These exact results were then used as benchmarks for the quantum hardware, PEPO simulations, and other classical approaches. It was found that PEPO techniques outperformed other methods in comparison to the exact results, achieving high accuracy with a bond dimension as small as $\chi = 2$. Moreover, at the Clifford point, the exact result was reproduced with $\chi = 2$, and the corresponding simulations required only about 3 seconds on a single CPU. For the deeper $n=20$ circuits, the results depended sensitively on the value of $\theta_h$. Consistent with the observations in \cite{ZalatelMPOIBM2023}, PEPO results were found to converge with increasing bond dimension and to agree with other methods for $\theta_h < \pi/8$. In contrast, for $\theta_h > 5\pi/16$, the PEPO results agreed with those obtained using Google 31 \cite{KechedzhiGoogle312024}, CPT \cite{ChanIBMClifford2023}, and MPO simulations, all of which converged to zero, whereas methods such as BP, MPS, and isoTN deviated from this value. A summary comparing the performance of the different classical simulation methods along with their strength and challenges for the kicked Ising experiment is presented in Table~\ref{tab_comparibm}.
\begin{table}[htbp]
\begin{center}
\small
\begin{tabular}{ |p{2cm}||p{4cm}|p{4cm}|p{4cm}|  }
 \hline
 TN method & Strength & Challenges & Performance\\
 \hline
 MPS   & tried $\&$ tested method, can be contracted exactly and efficiently for moderate bond dimensions & requires 2d mapping, causing inefficient long-range interactions; challenging for large 2d or complex geometries & fails at $n=5$ verifiable regime as $\theta_h$ increases; fails to reproduce exact results at the Clifford point and large $\theta_h$ for $n=6$\\
 \hline
 MPO & similar to MPS, well-suited for Clifford circuits, with unitary–conjugate cancellation simplifying calculations & similar to MPS, enlarged Hilbert space and therefore, may not be able to reach very  large bond dimensions & reproduce exact results for all $\theta_h$ including the Clifford points at $n=5$, good agreement with IBM processor at $n=6$. \\
 \hline
 iso TN & targets 2d systems, efficient contraction of the 2d networks due to isometric conditions & can only represent states obeying 1d area law, Moses move incur additional errors, lattice geometry did not match the IBM processor & fails at $n=5$ verifiable regime as $\theta_h$ grows, inaccessible for weight 17 operator and $n=6$, inaccurate at Clifford point $\theta_h = \pi /2$ \\
 \hline
 BP & directly targets 2d systems matching IBM processor geometry; avoids full tensor environment computation & requires Vidal gauging; breaks beyond tree-like correlations, with uncontrolled approximations & accurate and highly efficient simulation at $n=5$ and $n=6$ with respect to the exact results for all $\theta_h$ \\
 \hline
 g-PEPS & directly reflects processor geometry, without full environment computation or additional tensor gauging & expected to fail beyond tree-like correlations, with poorly controlled approximations & accurate, efficient simulation of $n=5$ and $n=6$ circuit for all values of $\theta_h$ matching exact results \\
 \hline
 PEPO & directly adapts 2D geometry, identifies low-rank entanglement, and exploits unitary–conjugate cancellation &enlarged Hilbert space limits bond dimensions; contraction may be inefficient at large $D$ & accurately and efficiently simulates $n=6$ across all 
$\theta_h$, lowest error compared to other techniques \\
 \hline
\end{tabular}
\caption{Table comparing the different classical Tensor network techniques used to benchmark IBM's 127 qubit Eagle processor in the Kicked Ising experiment~\cite{IBMkickedisingNat2023}.}
\label{tab_comparibm}
\end{center}
\end{table}

\newpage
\subsection{D-Wave}
D-Wave’s quantum computing hardware adopts an approach that differs fundamentally from most other quantum platforms~\cite{DWave}. Rather than implementing the gate-based circuit model, it is based on quantum annealing processors that employ thousands of superconducting qubits to address optimization and sampling problems by searching for the minimum-energy configuration of a programmable energy landscape. These qubits are realized using superconducting circuits operated at cryogenic temperatures and are connected through specialized lattice topologies, such as the Pegasus graph, which are designed to enhance connectivity and facilitate efficient problem embedding. D-Wave’s Advantage and Advantage2 systems have scaled to the level of several thousand qubits and are commercially available via both cloud-based and on-premises deployments. These systems target practical applications including materials modeling, combinatorial optimization in logistics, and certain machine learning tasks. In the following, we discuss a recent experiment that reports a claim of quantum advantage using D-Wave’s quantum annealing hardware~\cite{dwavesupremacy2025}.

\subsubsection{The quantum anealing experiment}
Recently, the D-Wave team reported an experiment based on a closely related quantum simulation problem and claimed computational supremacy over existing classical approaches~\cite{dwavesupremacy2025}. In this work, the authors simulated the dynamics of the Transverse Field Ising Model using superconducting quantum annealers on a variety of lattice geometries and graph structures. The time-dependent Hamiltonian is decomposed into two contributions as follows:
\begin{equation}
    H(t) = \Gamma (t/t_a) H_D + \mathcal{J}(t/t_a) H_P
\end{equation}
where the driving Hamiltonian is
\begin{equation}
    H_D = - \sum_i \sigma_i^x
\end{equation}
and the classical Ising part being
\begin{equation}
    H_P = - \sum_{i<j} J_{ij} \sigma_i^z \sigma_j^z
\end{equation}
The experiment begins with time evolution at $t=0$, where $\Gamma(0) \gg \mathcal{J}(0)$, placing the system deep in the paramagnetic phase. It ends at $t=t_a$, where $\Gamma(1) \ll \mathcal{J}(1)$, driving the system into the spin-glass phase due to the presence of random couplings. The quantum phase transition separating these two regimes is governed by the underlying topology of the experimental setup, which includes square, dimerized cubic, diamond, and dimerized biclique lattices. The experiments were carried out on D-Wave quantum annealing processors, namely Advantage1 (ADV1) and Advantage2 (ADV2), two hardware generations with distinct annealing schedules and comprising 5,627 and 1,222 qubits, respectively. We refer the reader to Ref.~\cite{dwavesupremacy2025} for detailed descriptions of the quantum processors, annealing protocols, and experimental implementation. The central task is to sample the quantum state obtained after a quench governed by the Hamiltonian above, for different annealing times $t_a$, and to compare the resulting observables with those obtained using classical simulation methods. To this end, the authors focus on 2D spin-glass instances defined on cylindrical $L \times L$ square lattices with system sizes up to $L=8$. These sizes are deliberately chosen to be sufficiently challenging for classical algorithms, while still allowing the computation of reliable ground-state reference data using Matrix Product States for benchmarking purposes. From the sampled states, two key quantities are evaluated: (a) the spin-glass order parameter, defined as
\begin{equation}
\langle q^2 \rangle = \frac{2}{N(N-1)}\sum_{i<j} c_{ij}^2,
\end{equation}
and (b) the residual energy, $E_{\text{res}} = \langle \mathcal{H}p\rangle - E_0$. Here, $c{ij}=\langle \sigma_i^z \sigma_j^z \rangle$ denotes the two-point spin correlation function, and $E_0$ is the ground-state energy of $\mathcal{H}_p$. The results obtained on the quantum processing units (QPUs) are benchmarked against MPS ground-truth calculations. Their level of agreement is quantified using the correlation error,
\begin{equation}
\epsilon_c = \left( \frac{\sum_{i,j}(c_{ij}-\tilde{c}{ij})^2}{\sum{i,j}\tilde{c}^2_{ij}}\right)^{1/2},
\end{equation}
where $\tilde{c}_{ij}$ denotes the ground-truth two-point correlation function. In addition to the ground-truth MPS simulations (with bond dimension $\tilde{\chi}=256$), further classical simulations were performed using Matrix Product States with smaller bond dimensions, Projected Entangled Pair States PEPS, and Neural Quantum States (NQS)~\cite{NQSreviewLange2024}. These classical results are systematically compared with both the QPU data and the MPS ground-truth values. We briefly discuss the resulting comparisons below.

\subsubsection{Tensor network simulations of D-Wave experiment}
For a $6 \times 6$ square lattice with cylindrical boundary conditions, the results obtained from the QPU were directly compared with those from MPS simulations. Across annealing times spanning three orders of magnitude, excellent agreement was observed between the QPU data and the MPS ground-truth results with bond dimension $\chi=256$, both for the squared spin-glass order parameter $\langle q^2\rangle$ and for the residual energy per spin, $E_{\text{res}}/N$. Focusing on the same system size and an annealing time of $t_a = 7\mathrm{ns}$, the authors further examined the spin–spin correlation functions $c_{ij}$ by comparing them with the corresponding ground-truth values. They found that reproducing results of comparable quality to those obtained from the QPU required MPS simulations with a significantly smaller bond dimension, $\chi=64$, yielding correlation errors of $\epsilon_c = 0.059$ and $\epsilon_c = 0.081$, respectively. In addition, an analysis based on the individual state probabilities revealed similarly close agreement between the QPU and MPS results at $\chi=64$, as quantified by state fidelities of $\mathcal{F} = 0.752$ and $\mathcal{F} = 0.746$, respectively.

For 20 disorder realizations on general $L \times L$ lattices, it was observed that the median correlation error $\epsilon_c$ remains nearly constant across all system sizes when using the QPU. In contrast, achieving a comparable error level with MPS simulations requires an exponential increase in the bond dimension, denoted $\chi_Q$, for each of the two annealing times considered. Based on these observations, the authors conclude that, given their available computational resources, MPS simulations are able to match or surpass the QPU performance only for system sizes up to $L=8$ and across all annealing times studied. 

The authors conclude that both PEPS and NQS methods encounter significant difficulties even for relatively small system sizes when considering slower quenches ($t_a = 20,\mathrm{ns}$). In the case of PEPS, the simulations were performed using the simple-update scheme, with the size of the local update neighborhood progressively increased until it spanned the full $8 \times 8$ lattice. This approach, however, led to rapidly growing numerical costs. Even the best-performing PEPS variant, referred to as Plqt-NN+, failed to outperform the QPU in terms of correlation error for the slow-quench case of $t_a = 20,\mathrm{ns}$. Moreover, increasing the PEPS bond dimension $D$ did not lead to a noticeable improvement in accuracy. The authors attribute the limited performance of PEPS to the emergence of long-range, system-spanning correlations during slower annealing protocols. A qualitatively similar conclusion is reached for Neural Quantum States, which also fail to achieve comparable accuracy under these conditions. On this basis, the authors argue that neither PEPS nor NQS provides a competitive classical baseline for benchmarking the QPU in this regime, and they therefore restrict their classical comparisons to MPS simulations when considering other lattice topologies.

For cubic and diamond lattices, the authors report that the QPU error remains approximately constant as a function of system size, in close analogy with the behavior observed for square lattices. In contrast, for biclique lattices the error increases with system size. On the classical side, the MPS bond dimension $\chi_Q$ required to match the quality of the QPU results is found to grow exponentially with system size, consistent with scaling expectations based on the bipartition area. Interestingly, for all lattice topologies considered, $\chi_Q$ decreases with increasing annealing time $t_a$. Using the observed scaling of the required MPS bond dimension $\chi_Q$ with system size, the authors estimate the classical computational resources needed to reproduce the QPU results. They conclude that, for the largest instances accessible to the QPU, matching the reported accuracy with MPS would require on the order of millions of years of computation on the Frontier supercomputer, together with approximately $700\mathrm{PB}$ of storage and an electricity consumption comparable to the current annual global usage. Finally, to assess the reliability of the QPU results beyond the classically tractable regime, the authors invoke concepts from the theory of critical phenomena. In particular, they extract the Binder cumulant and the Kibble–Zurek (KZ) exponent, finding values in good agreement with theoretical expectations. These results formed the basis of the claims of quantum advantage by D-Wave.

In Ref.~\cite{Tindalldisordyn2d3d}, the authors reported accurate and efficient classical simulations of the quantum annealing problem described above using belief-propagation methods formulated within two- and three-dimensional tensor network frameworks. For certain classes of systems, they demonstrated that the annealing dynamics can be simulated efficiently on contemporary classical computers, with computational resources scaling only linearly with system size for a fixed annealing time. Using these classical approaches, the authors achieved accuracies that surpassed those of the quantum hardware for square lattices with cylindrical boundary conditions as well as for diamond lattices, while attaining comparable accuracy for dimerized cubic lattices. Moreover, by simulating systems comprising more than 300 qubits, they were able to extract the Kibble–Zurek exponent associated with crossing the quantum critical point, obtaining values consistent with those reported in the existing literature. As demonstrated in Ref.~\cite{TindallBPPRX2024}, the authors employ a tensor network ansatz that explicitly reflects the geometry of the underlying lattice—an essential and nontrivial consideration for efficiently capturing the relevant correlations. Their algorithm is based on a simple-update scheme, relying on standard belief propagation for time evolution, and employs different BP-based variants for the evaluation of observables. The first variant is an adaptation of the boundary Matrix Product State method~\cite{LubaschPEPSPRB2014,VerstraetePEPS2004}, in which MPS are used to approximate the contraction of 2D tensor networks in a layer-by-layer fashion. This approach is also commonly referred to as MPS message passing~\cite{BPGuo2023}. The second variant is the so-called loop-corrected BP method, which expresses the tensor network contraction as a sum over configurations involving projectors and anti-projectors constructed from the message tensors~\cite{EvenblyLoopBP2024}.

For the $8 \times 8$ square lattice with cylindrical boundary conditions, the authors demonstrate that using standard BP for time evolution together with message-passing BP for the evaluation of observables yields errors that are lower than those obtained with PEPS in the original experiment, as well as lower than those produced by the quantum annealer, even for long annealing times of $t_a = 20,\mathrm{ns}$. Moreover, they find that, for both annealing times considered ($t_a = 7,\mathrm{ns}$ and $t_a = 20,\mathrm{ns}$), the error decreases exponentially with increasing cylinder circumference (and hence with the number of qubits) when observables are computed using the message-passing BP technique, ultimately falling below the corresponding errors reported for the quantum annealer. For sufficiently large cylinders, message-passing BP consistently outperforms the loop-corrected BP approach. For this lattice geometry, the simulations were carried out with a bond dimension of up to $\chi_{\mathrm{BP}} = 32$ during the time-evolution stage (i.e., during gate application). For the evaluation of observables, the final state was truncated to $\chi = 10$, and message tensors with bond dimension $r = 2\chi$ were employed. For cubic and diamond lattices, the authors instead used loop-corrected BP to compute observables, finding lower errors than the quantum annealer for the diamond lattice and comparable errors for the cubic lattice. Importantly, both approaches for computing observables—message-passing BP and loop-corrected BP—exhibit computational costs that scale linearly with the number of qubits. As a result, these methods are significantly more efficient than the MPS simulations used to obtain ground-truth benchmarks. For example, to achieve converged results for a three-dimensional diamond lattice comprising 50 qubits, MPS simulations required several days of runtime on an Intel Skylake CPU, whereas the BP-based methods achieved convergence within approximately one hour on the same hardware. We summarize the performance of the different methods in Table~\ref{tab_compardwave}.

\begin{table}[htbp]
\begin{center}
\small
\begin{tabular}{ |p{2cm}||p{4cm}||p{4cm}||p{4cm}|}
 \hline
 TN method & Strength & Challenges & Performance\\
 \hline
 MPS   & tried $\&$ tested method, can be contracted exactly and efficiently for moderate bond dimensions & requires 2d mapping, causing inefficient long-range interactions; challenging for large or complex geometries and higher dimensions. & matches or exceeds QPU performance on square lattices up to $8 \times 8$ for all annealing times; beyond this, $\chi_Q$ grows exponentially with system size, while decreasing with longer annealing times (7–20 ns).\\
 \hline
 PEPS & directly targets 2d lattices, does not require additional gauging of tensors & System spanning correlations, results did not improve with increasing bond dimensions & performance depends on annealing time $t_a$: at $2$ns, PEPS yields lower $\epsilon_c$
 than the QPU with moderate $D$; for larger $t_a$, performance saturates despite increasing $D$.\\
 \hline
 PEPS + BP & directly targets 2d lattices, algorithm complexity scales linearly with  the number of  qubits & did not attempt complex lattice geometry, large simulations in 3d lattices, longest simulation times, etc & outperforms the QPU for all annealing times on 8 $\times$ 8 cylindrical and 3d diamond lattices; comparable on cubic lattices.\\
 \hline
\end{tabular}
\caption{Table comparing the different classical Tensor network techniques used to benchmark D-Wave's superconducting quantum processor in their annealing experiment~\cite{dwavesupremacy2025}.}
\label{tab_compardwave}
\end{center}
\end{table}

\newpage
\subsection{Google}
Google’s quantum computing hardware is based on superconducting qubit processors developed to advance scalable, high-performance quantum computation~\cite{Google}. The company’s latest flagship processor, Willow, employs a 2D grid of superconducting transmon qubits, enabling high-fidelity gate operations and improved error-correction capabilities—both critical milestones on the path toward fault-tolerant quantum computing. Earlier generations, most notably the Sycamore processor, established key benchmarks in quantum information processing, including experiments that reported “quantum supremacy” for carefully selected computational tasks. Google complements its hardware with custom cryogenic control electronics and a dedicated software stack, including the Cirq framework~\cite{Cirq}, while continuously improving qubit coherence, connectivity, and error-mitigation strategies. These developments are aimed at enabling increasingly complex quantum simulations and algorithms that may outperform classical approaches for specific, well-defined problems. Below, we review two quantum-advantage claims reported using Google’s superconducting quantum processors.

\subsubsection{Random Circuit Sampling}
One of the earliest experiments to claim quantum supremacy (or quantum advantage) using Random Circuit Sampling (RCS) was reported in 2019 by Google’s team~\cite{Googlequantumsupremacy2019}. Random Circuit Sampling tasks a quantum processor with generating bitstrings drawn from the output distribution of an $n$-qubit quantum circuit of depth $d$. These circuits are deliberately constructed to frustrate efficient classical simulation. In particular, circuits with low entanglement, Clifford-only gates, or strongly peaked output distributions admit efficient classical algorithms and are therefore avoided. Instead, RCS employs randomly chosen single-qubit gates interleaved with entangling two-qubit gates arranged according to a 2D connectivity pattern. Each application of single- and two-qubit gates constitutes a circuit layer, and stacking many such layers rapidly generates highly entangled quantum states. When the circuit is initialized in the all-zero state, its execution defines a probability distribution over output bitstrings. For sufficiently deep and well-designed random circuits, this distribution is expected to follow the Porter–Thomas form, a hallmark of chaotic quantum dynamics~\cite{boixoNature2018, Googlequantumsupremacy2019}.

Because real quantum hardware is inherently noisy, experimental demonstrations must quantify how closely the device’s output distribution, $p_{\mathrm{device}}(x)$, approximates the ideal distribution $p_{\mathrm{ideal}}(x)$ for bitstrings $x$. This comparison is commonly performed using the linear cross-entropy benchmarking (XEB) fidelity,
\begin{equation}
\label{eq:xeb} 
F_{\mathrm{XEB}} = 2^n \, \mathbb{E}_{x \sim p_{\mathrm{device}}} \big[ p_{\mathrm{ideal}}(x) \big] - 1, 
\end{equation}
where the value of $p_{\mathrm{ideal}}(x)$ for each sampled bitstring is obtained by classically simulating the corresponding quantum circuit. In the absence of noise, XEB approaches unity. In particular, if the device produces samples exactly according to the ideal distribution, then $p_{\mathrm{device}} = p_{\mathrm{ideal}}$. For a Haar-random quantum state, the ideal output distribution follows the Porter–Thomas form. For such a distribution, one can show that $2^n \, \mathbb{E}_{x \sim p_{\mathrm{ideal}}}
      \big[ p_{\mathrm{ideal}}(x) \big]
    = 2.$ Substituting into the definition above gives $F_{\mathrm{XEB}} = 1.$

Intuitively, for a uniformly random quantum state, most bitstrings occur with probabilities $p_{\mathrm{ideal}}(x) \approx 2^{-n}$, while a small fraction appear with significantly larger probabilities. If a device samples from the correct distribution, the observed bitstrings are therefore biased toward outcomes with slightly higher-than-average values of $p_{\mathrm{ideal}}(x)$. This positive correlation leads to a scaled expectation value $2^n,\mathbb{E}[p_{\mathrm{ideal}}(x)] = 2$, and consequently to $F_{\mathrm{XEB}} = 1$. If, on the other hand, the device output is partially mixed with uniform noise,
\begin{equation}
p_{\mathrm{device}} = (1-\epsilon) p_{\mathrm{ideal}}
                       + \epsilon \cdot 2^{-n},
\end{equation}
then $F_{\mathrm{XEB}} = 1 - \epsilon,$ which means the XEB value directly estimates the circuit fidelity. For the largest and most complex circuits (53 qubits, 20 cycles), Google's team estimated the fidelity to be $F_{\mathrm{XEB}} = (2.24 \pm 0.21)\times10^{-3}$, allowing the authors to assert with $5\sigma$ confidence that the average fidelity is greater than at least $0.1\%$.

\textbf{Google's classical benchmarks:} Classical simulation played a dual role in the experiment. It was used both to verify the device output for simpler circuits and to estimate the computational cost of simulating the most challenging instances. For circuits involving up to 43 qubits, the authors performed exact simulations of the full quantum state. The largest of these simulations were executed on the Jülich supercomputer, utilizing approximately 100,000 CPU cores and 250 terabytes of memory. For circuits exceeding 43 qubits, the simulation strategy was modified by partitioning the circuit into two subcircuits (or patches), each of which was simulated exactly. The results were then recombined using a method inspired by Feynman path integrals. While this hybrid approach significantly reduced the computational requirements for moderate circuit depths, its cost grows exponentially with increasing depth. To extend the reach of these simulations, the authors further optimized the path-integral contraction for low-depth circuits and executed these calculations on the Summit supercomputer.

To estimate the classical computational cost of simulating the full supremacy circuits, the team executed portions of the simulation on both the Summit supercomputer and Google’s internal computing clusters, and then extrapolated to the total cost. This extrapolation explicitly accounted for the reduced computational burden arising from the low output fidelity $F_{\mathrm{XEB}}$ of the quantum processor; for instance, an average fidelity of $0.1\%$ reduces the classical simulation cost by approximately a factor of $10^3$. For the largest circuit considered ($n=53$, $m=20$) with an estimated fidelity of $0.1\%$, the authors estimated that simulating the circuit using the hybrid “Method 2” approach on Google Cloud servers would require roughly $5\times10^{13}$ core-hours and consume on the order of one petawatt-hour of energy. Moreover, classical verification of the quantum output—namely, the computation of the probabilities $P(x_i)$ required to evaluate $F_{\mathrm{XEB}}$ for the most complex circuits—was projected to take millions of years. These classical benchmarks further highlighted the separation in performance: for circuits with $m=20$ cycles, obtaining one million samples on the quantum processor required approximately 200 seconds, whereas generating samples of comparable fidelity using classical methods was estimated to take on the order of $10^4$ years on a classical system employing one million CPU cores.

\subsubsection{Tensor network simulations of Google's RCS experiment}
\begin{figure}
	\begin{center}
		\includegraphics[width=\textwidth]{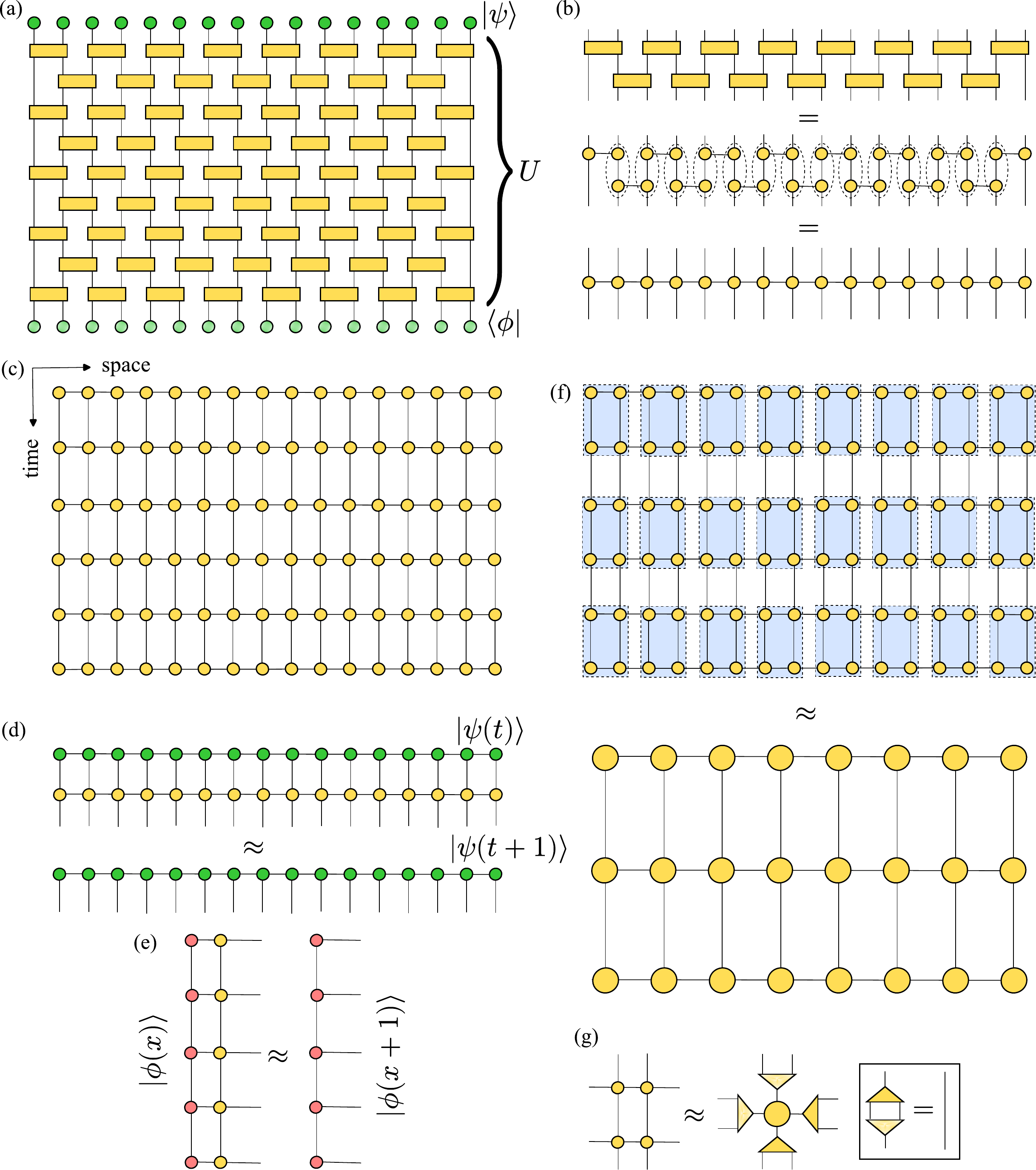}
		\caption{Random circuit sampling via tensor networks. (a) The tensor network representing the probability amplitude for a product state $|{\phi}\rangle$ in the evolved quantum state $U|\psi\rangle$ where $U$ is expressed as a brickwall circuit of (possibly, random) two-body gates. (b) Each layer of circuit can be regrouped as an MPO. (c) This yields a square-lattice tensor network whose horizontal and vertical directions correspond to space and time, respectively. The resulting network admits multiple contraction strategies, including boundary-MPS propagation along (d) time or (e) space, as well as (f–g) coarse-graining–based contraction methods.
        }
    \label{fig:rcs_contract}
	\end{center}
\end{figure}

From a tensor network perspective, estimating the probability distribution generated by a (random) quantum circuit requires repeatedly contracting a tensor network corresponding to $\langle{\psi}|\rm{circuit}|{\phi}\rangle$; see Fig.~\ref{fig:rcs_contract}. The first well-founded challenge to Google’s experiment appeared in Ref.~\cite{Gray2021hyperoptimized}. In this work, the authors introduced new tensor network–based algorithms that substantially improved contraction-path discovery and estimated a speedup of approximately $10^4$ relative to Google’s original 10,000-year classical runtime estimate. By combining hypergraph partitioning, slicing techniques, and highly optimized contraction trees, they projected that the Summit supercomputer could reproduce the experiment in roughly 195 days at an effective utilization of $68\%$ of peak FLOPS. A central contribution of this work is the introduction of hyper-optimized tensor network contractions. Unlike earlier approaches that focused primarily on minimizing the leading asymptotic scaling—typically characterized by the maximum contraction width $W$—this framework aims to optimize the entire contraction tree with respect to the total contraction cost $C$. Here, $C$ quantifies the overall time complexity in terms of the total number of scalar operations, while $W$ captures the space complexity through the size of the largest intermediate tensor. 

The framework employs randomized optimization strategies enhanced by stochastic Bayesian optimization to efficiently explore and refine contraction paths. Contraction trees are constructed explicitly by combining agglomerative (bottom-up, e.g., Hyper-Greedy), divisive (top-down, e.g., Hyper-Par), and optimal drivers for building subtrees. A crucial preprocessing stage consists of a series of efficient local searches, referred to as simplifications, which are applied iteratively to reduce the complexity of the tensor network prior to the main path-finding procedure. These simplifications include: (1) \textit{Diagonal Reduction} -- replaces a tensor with a lower-dimensional tensor when its entries vanish for unequal index pairs. This transformation can introduce hyperedges, thereby enabling the use of hypergraph-based methods. (2) \textit{Rank Simplification} -- absorbs rank-1 and rank-2 tensors into neighboring tensors to reduce the overall rank of the network. (3) \textit{Anti-diagonal Gauging} -- permutes the ordering of indices to transform tensors into a form amenable to diagonal reduction. (4) \textit{Column Reduction} -- removes an index entirely when the tensor is nonzero only for a single value of that index, effectively projecting the index onto a fixed basis state. (5) \textit{Split Simplification} -- performs an exact low-rank decomposition via singular value decomposition (SVD) across a bipartition of indices, reducing the cut weight even at the expense of increasing the total number of tensors.

In Ref.~\cite{StoudenmireGoogleRCS}, the authors exploited the presence of noise to simplify classical simulation. Specifically, they employed time-evolving block decimation (TEBD) with controlled truncation to emulate fidelity loss, and simulated 54-qubit random circuits of depth 20—albeit using CZ gates rather than Google’s higher-rank $\sqrt{\mathrm{iSWAP}}$ gates. Using a maximum MPS bond dimension of 320, they were able to reach fidelities comparable to those reported by Google in under 48 hours on a single CPU core. However, because the simulated circuits differ substantially from Google’s experimental circuits, and because the notion of fidelity employed in this approach is not directly equivalent to the XEB fidelity used in the experiment, these results are not regarded as a direct refutation of the original quantum-advantage claim.

In Ref.~\cite{huang2020classicalsimulationquantumsupremacy}, the authors introduced a further tensor network–based approach that improves contraction-path optimization. For random circuits with 53 qubits and depth 14, they reported the generation of three million samples at approximately $1\%$ fidelity in 264 seconds—about a factor of two faster than Google’s Sycamore processor. By extrapolating their results to depth 20, they estimated a runtime of roughly 12 hours to achieve a fidelity of $0.2\%$. However, this extrapolation relied on unrealistically high floating-point efficiency and was not validated by explicit simulations. Consequently, this work is not regarded as a decisive challenge to Google’s 2019 quantum-supremacy experiment. Methodologically, the approach focuses on further optimizing tensor network contractions through a combination of techniques. The central idea is to identify the high-cost “stem” of the contraction tree and optimize it using strategies such as hypergraph partitioning and dynamic slicing, thereby reducing the overall contraction cost. As in earlier tensor network–based simulations, the method keeps a fixed number of time-like edges (six in this case) open in order to compute many output amplitudes simultaneously, and subsequently employs frugal rejection sampling to efficiently generate bitstrings from the resulting probability distribution.

In Ref.~\cite{pan2021simulatingsycamorequantumsupremacy}, researchers from the Chinese Academy of Sciences introduced the so-called big-head algorithm for classically simulating Google’s random-circuit-sampling experiment. By selecting a large number of qubits to remain open and carefully optimizing the contraction structure, they were able to sample one million correlated bitstrings from 53-qubit, depth-20 circuits in approximately five days using large GPU clusters. After post-processing, the authors reported an XEB fidelity of $0.739$, substantially higher than that achieved in Google’s experiment, effectively reducing the original 10,000-year classical runtime estimate to about five days—albeit with the important caveat that the generated samples are correlated. Conceptually, this work extends earlier tensor network simulation approaches by keeping a much larger number of qubits open in order to compute larger batches of amplitudes simultaneously, in contrast to previous methods that typically left only a small number (e.g., six) of qubits open. While opening additional qubits greatly increases the computational complexity of the contraction, it also enables significantly higher throughput. Selecting the optimal subset of open qubits constitutes a challenging combinatorial optimization problem. To address this challenge, the authors first determine an efficient contraction order for the full tensor network and subsequently choose which qubits to keep open based on this ordering. The resulting contraction tree is divided into two components: a head, which contains the open qubits, and a tail, which contains the closed qubits. The head and tail sub-networks are then partitioned and contracted independently, and their results are combined via a dot product. This strategy—designed to maximize the number of open qubits in the head while controlling contraction cost—is referred to as the big-head algorithm. In addition, dynamic slicing is employed to further accelerate the contraction of both sub-networks, building upon techniques developed in earlier tensor network–based simulations.

In Ref.~\cite{LiuGoogleRCSSupremacy}, a large-scale simulation performed on the Sunway supercomputer demonstrated that a single bitstring from circuits inspired by Google’s experiment could be sampled in 304 seconds using a PEPS-based tensor network contraction algorithm specifically optimized for the underlying hardware architecture. Although the circuit definitions in this study differed from those used in Google’s experiment, the results indicated substantial additional classical performance gains. Subsequent projections suggested that the runtime could potentially be reduced further to 60.4 seconds under optimized conditions. In their largest reported simulations, the authors utilized 41,932,800 Sunway cores, each equipped with 32 GB of memory and a peak performance of 4.7 TFlops. They considered random circuits with up to 100 qubits and 42 cycles. However, these circuits differed from Google’s experimental circuits in both the definition of circuit depth and the choice of two-qubit gate, factors that significantly influence classical simulation complexity. Shortly after, the Sunway team reported a further milestone by applying the big-head method to circuits that matched Google’s 2019 experimental definitions exactly, including qubit count, cycle depth, two-qubit gate set, and overall circuit structure. Using more than 41 million cores, they sampled one uncorrelated bitstring in 400 seconds using single precision and in 276 seconds using mixed precision arithmetic. They also demonstrated significantly shorter runtimes for shallower circuits. This implementation combines the big-head algorithm with earlier Sunway-specific optimization strategies. In this setup, six qubit wires are left open during contraction, allowing the simultaneous computation of 64 amplitudes per contraction, followed by frugal sampling to generate uncorrelated bitstrings. In addition to the 20-cycle circuits, the authors simulated 12- and 14-cycle circuits, producing samples in 18 and 82 seconds, respectively.

In Ref.~\cite{PanGoogleRCSSupremacy}, published two years after Google’s original announcement, the authors presented what has been described as a partial or “weak” refutation of the 2019 quantum-supremacy claim. Building upon their earlier big-head algorithm, they generated one million uncorrelated bitstrings from circuits matching the Sycamore architecture, achieving an XEB fidelity of approximately $0.0037$ in about fifteen hours using 512 GPUs. Although this runtime remains longer than Sycamore’s reported 200-second sampling time, the authors argued that, on a sufficiently large exascale computing cluster, the simulation time could be reduced to a few dozen seconds—potentially comparable to or faster than the quantum processor. For this reason, the result has been characterized as a weak refutation rather than a definitive one. The central idea of the approach is to approximate the tensor network by selectively removing certain edges. Breaking an edge is equivalent to inserting a Pauli-$Z$ projector, $(I+Z)/2$, which effectively replaces the corresponding local state with $|0\rangle$. Each removed edge reduces the fidelity by a factor of two; thus, removing eight edges yields a fidelity of $2^{-8} \approx 0.0039$, consistent with the reported value of $0.0037$ once additional minor approximations are taken into account. The resulting approximate tensor network is then contracted using the big-head algorithm combined with dynamic slicing. Six qubits are kept open during contraction to enable batch amplitude computation, and importance sampling is employed to eliminate correlations between samples. In total, approximately $2^{20}$ uncorrelated samples were generated in just over fifteen hours.

In 2024, Ref.~\cite{zhao2024leapfroggingsycamoreharnessing1432} reported what is widely regarded as the first strong classical refutation of Google’s 2019 quantum-supremacy experiment. Using a cluster of 1,432 GPUs, the authors generated three million samples in 86.4 seconds—surpassing Sycamore’s reported runtime of 200 seconds for one million samples. Their approach integrates several previously developed techniques, including the big-head algorithm, dynamic slicing, and optimized contraction-tree construction, together with Markov chain Monte Carlo (MCMC) sampling and post-selection to achieve an XEB fidelity of approximately 0.002, comparable to Sycamore’s reported value. In this implementation, ten qubits are kept open in the head tensor network, significantly increasing the number of amplitudes computed per contraction. Furthermore, due to careful slicing optimization, only about $0.03\%$ of the $2^{24}$ potential subtasks require explicit contraction. The resulting workload, distributed efficiently across the GPU cluster, completes in 86.4 seconds. Given that the classical runtime undercuts the quantum processor’s sampling time at comparable fidelity, this result is broadly viewed as the first robust classical challenge to Google’s 2019 quantum-supremacy claim.

\subsubsection{Google's quantum echoes experiments}
Most recently, the team at Google Quantum AI reported a new claim of quantum advantage using their superconducting Willow processor~\cite{Google_quantum_echo2025}. Referred to as the quantum echoes experiment, the work focuses on measuring higher-order, out-of-time-order correlators (OTOCs), which are believed to be intractable for classical simulation at the relevant system sizes and circuit depths. The authors further argue that, unlike random circuit sampling experiments, their results can in principle be independently verified using other quantum processors and may have potential relevance for real-world applications. The experiment was performed on Google’s Willow chip, a 2D array of 103 superconducting transmon qubits. The reported average single-qubit gate error is approximately $0.05\%$, while the two-qubit gate error is around $0.15\%$. The qubits are arranged on a square lattice, and a target qubit $q_m$ is measured in a Pauli basis $M \in {X, Y, Z}$. The time-ordered correlator (TOC), corresponding to a measurement at time $t$, is defined as $\langle M(t) M \rangle$, where $M(t) = U(t)^{\dagger} M U(t)$ denotes the Heisenberg-picture evolution of the operator $M$. In ergodic systems, the TOC typically exhibits exponential decay in time due to scrambling dynamics. In the experiment, however, the unitary evolution $U(t)$ is replaced by a nested echo sequence defined as $U_k(t) = B(t)[MB(t)]^{k-1}$, where $B(t)$ is another time-evolved Pauli operator acting on a qubit $q_b$ located some distance away, and $k \geq 1$ is an integer. The corresponding out-of-time-order correlator (OTOC) can be written as
\begin{equation}
    \mathcal{C}^{2k}= \langle U_k(t)^{\dagger} M U_k(t) M \rangle = \langle (B(t) M)^{2k} \rangle
\end{equation}

The authors then analyze the sensitivity of the OTOC to the microscopic details of the quantum dynamics, implemented as quantum circuits composed of random single-qubit gates and fixed two-qubit gates. A given circuit instance $i$ is generated by varying the random single-qubit gates while keeping the two-qubit gate pattern fixed. For each instance, the OTOC is measured repeatedly until statistical noise is reduced below a specified threshold, for a fixed number of circuit samples at evolution time $t$ under $U$. The experiment is repeated for different values of $t$, for different choices of measured qubits $q_m$ and $q_b$, and across multiple circuit instances $i$, with each configuration sampled between 50 and 250 times. One of the central findings is that the OTOC exhibits algebraic decay in time, with its standard deviation remaining above $0.01$ even beyond $t=20$. In contrast, the corresponding time-ordered correlator (TOC) decays exponentially, and its standard deviation drops below $0.01$ already by $t=9$. These results indicate that higher-order OTOCs retain sensitivity to microscopic circuit details even in regimes where the system is otherwise ergodic, highlighting their potential as probes of fine-grained scrambling dynamics.

The authors claim that classical techniques based on Monte Carlo sampling and tensor network methods are unable to reproduce higher-order OTOCs with a comparable signal-to-noise ratio (SNR). For 40-qubit circuits, the quantum processor achieved an SNR of 5.4, whereas classical methods reached 5.3 for diagonal components but only 1.1 for off-diagonal components. For larger circuits involving 65 qubits, classical simulation on the Frontier supercomputer was estimated to require approximately 3.2 years per circuit, compared with $2.1$ hours for the corresponding experimental measurement—placing the experiment, according to the authors, beyond the frontier of classical tractability. To further demonstrate the potential practical relevance of higher-order OTOCs, the authors considered an application to Hamiltonian learning. Specifically, they studied a physical system governed by a Hamiltonian containing unknown parameters and generated a dataset of second-order OTOCs. These data were compared against quantum simulations of the same Hamiltonian, and the unknown parameters were iteratively optimized until agreement with the measured values was achieved. In the reported implementation, a 34-qubit random circuit was used, with the unknown parameter being a phase $\zeta$ in one of the two-qubit gates. The authors generated 20 distinct circuit instances (each differing in their local single-qubit gate choices) and treated their classically simulated OTOC(2) values as surrogate “experimental” data for training. They then measured the off-diagonal components of OTOC(2), denoted $C^{(4)}_{\mathrm{off\text{-}diag}}$, on the quantum processor across a range of trial values of $\zeta$. For all circuit instances considered, the experimentally measured data intersected the ideal (classically simulated) curves precisely at the true target value of $\zeta$. From this observation, the authors concluded that second-order OTOCs encode sufficiently detailed information about the system dynamics to enable the recovery of unknown Hamiltonian parameters.

\subsubsection{Tensor network simulations of Google's quantum echoes experiments}
As in the earlier experiments discussed above, the authors employ tensor network–based techniques as one of the primary classical benchmarking tools. They map the exact computation of expectation values, $\langle Z \rangle = \langle 0 |C Z C^{\dagger} |0 \rangle$ for a quantum circuit $C$, to a tensor network contraction (TNC) problem. However, exact contraction of such networks becomes computationally intractable for large numbers of qubits and deep circuits. To mitigate this difficulty, various approximation strategies can be employed. In this work, the authors rewrite the expectation value as a sum over orthogonal projections applied to all qubits at the measurement layer, excluding the measured qubit and a small set of nearby gates:
\begin{equation}
\begin{split}
    \langle Z \rangle &= \langle 0 | C Z \left[ \sum_x \Pi_x\right]C^{\dagger}|0 \rangle\\
    &= \sum_x \langle \psi_x|Z|\psi_x \rangle
\end{split}
\end{equation}
where $|\psi_x \rangle$ denotes a partial quantum state conditioned on a specific projection $x$. Instead of summing over all possible projections $x$, an approximate contraction of the tensor network can be obtained by sampling only a subset of projections. The authors implement this strategy using rejection sampling: random projections are drawn, and the expectation value is estimated as
\begin{equation}
    \langle Z \rangle = \mathbb{E} \left[ \frac{\langle \psi_x|Z|\psi_x \rangle}{p_x}\right]
\end{equation}
where $p_x = |\langle \psi_x|\psi_x \rangle|^2$ represents the corresponding probability weight. In practice, the rejection sampling proceeds by exactly computing $p_x$ for a small number of randomly selected projections drawn uniformly from ${x_1,x_2,...,x_k}$. Each candidate projection is then accepted or rejected with probability $r_x = \frac{p_x} {max\{x_1 x_2...x_k\}p_x}$ ensuring unbiased sampling of higher-weight contributions.

After introducing the sampling procedure described above, the authors developed a Tensor Network Contraction Optimizer (TNCO) to compute the resulting approximate expectation values. For small circuits, where exact simulation of the OTOC remains feasible, they observed a clear correlation between the OTOC values obtained using tensor network Monte Carlo (TNMC) techniques and the exact results, typically accompanied by relatively large SNRs. The authors then evaluated the performance of TNMC on both small circuits and larger circuits, for which exact simulation is no longer tractable, by comparing the classical results with experimental OTOC measurements. They observed a degradation of the SNR when benchmarking against experimental data. Moreover, the SNR was found to decrease with increasing circuit size, while improving as the number of tensor network cuts was reduced. In addition to TNMC, several other classical simulation techniques were investigated, including Clifford expansion methods, Pauli-weight truncation algorithms, and variational approaches such as neural quantum states and matrix product states. For the MPS simulations, the 2D circuit was mapped onto a 1D tensor network using a snake-like ordering. The analysis focused on two approximation strategies: (i) truncating Schmidt values without removing gates, and (ii) removing selected gates without performing Schmidt truncation. Both approximations affect the achievable SNR but are necessary to control the rapid growth of bond dimension resulting from the strong entanglement generated by the circuits. Matrix Product Operator techniques were also employed to evolve operators directly. However, these simulations were found to be more computationally demanding than direct state-vector simulations. Overall, the authors concluded that none of the alternative classical approaches achieved a competitive SNR in simulating OTOCs compared to the tensor network contraction strategy. On this basis, they inferred that optimized tensor network contraction currently represents the most efficient classical method for OTOC simulation in this regime.

\section{Directions for Quantum Advantage}
\label{sec:lims_directions}
The quantum-advantage experiments discussed above, together with the subsequent tensor network–based classical benchmarks, raise several important and timely questions regarding the limits of classical simulation and the future direction of quantum computational advantage. While IBM’s Kicked Ising experiment did not ultimately establish a clear quantum advantage, it nonetheless generated substantial progress with broad impact on the field. First, it stimulated the development of new classical simulation methods, providing valuable insight into their capabilities, scaling behavior, and limitations. Such advances have significantly sharpened our understanding of where tensor network approaches succeed and where they begin to break down. Second, these efforts have highlighted the importance of identifying problem instances and circuit constructions that genuinely challenge the structural assumptions exploited by classical algorithms. In particular, they suggest that merely increasing system size or circuit depth is insufficient; rather, one must design dynamics that fundamentally frustrate efficient contraction, low-entanglement approximations, or favorable graph decompositions. Third, these developments indicate that demonstrating quantum advantage in a convincing manner may require benchmarking not only against state-of-the-art classical methods but also across multiple quantum hardware platforms. Focusing on regimes where classical methods are expected to fail, while simultaneously validating results through independent quantum processors, could provide a more robust and platform-independent notion of quantum advantage. In the following, we examine the scope and limitations of leading tensor network approaches and analyze why certain methods fail to efficiently simulate particular quantum experiments. We also highlight the risks associated with selecting an inappropriate tensor network ansatz as a benchmark, which can lead to misplaced confidence in claims of quantum advantage. Conversely, we discuss why specific tensor network strategies have successfully simulated certain quantum hardware experiments, delineate the boundaries of their applicability, and explore the implications of these findings for future claims of quantum advantage.

\subsection{Scope and Limits of Tensor Networks for Quantum Advantage}
A particular thing to keep in mind is the fundamental limitation of tensor network methods, namely, entanglement capacity: area-law versus post–area-law scaling, and the geometry of correlations encoded by a given ansatz. However, entanglement-scaling arguments describe asymptotic behavior, whereas supremacy demonstrations typically involve on the order of one to two hundred qubits. At these intermediate scales, constant prefactors and implementation details dominate, and entanglement exponents alone lose predictive power. The relevant distinctions between tensor network ansätze are governed less by their formal expressiveness and more by practical constraints: the maximum bond dimension that can be reached, the numerical stability of the optimization procedure, and the computational cost of contraction. 

This tension is most clearly illustrated by the simplest tensor network ansatz: the \emph{MPS}. A fixed-bond-dimension MPS has well-understood entanglement limitations; for example, a random MPS saturates its bipartite entropy bound and exhibits exponentially decaying correlations. Yet the structural simplicity of the MPS allows simulations with exceptionally large bond dimensions. Recent large-scale DMRG implementations running on 1024 TPU v3 cores have reached $\chi = 2^{16} = 65{,}536$, sufficient to represent arbitrary 32-qubit states arranged on a 1D chain exactly~\cite{DMRGtpu}. This benchmark effectively delineates the current practical frontier of MPS-based classical simulation. Yet even MPS becomes unpredictable in the context of circuit simulation. The effective bond dimension depends sensitively on the chosen contraction boundary. Propagating along the circuit’s time direction can generate rapid entanglement growth and correspondingly large intermediate bond dimensions. In contrast, contracting along a spatial direction—where indices are naturally time-ordered—may be significantly more manageable, depending on the decay of temporal correlations. Thus, the same quantum circuit may be either tractable or intractable depending on how the tensor network is sliced. While bond dimension remains the central bottleneck, the geometry of the contraction boundary determines how rapidly that bottleneck is encountered. 

Additionally, MPS techniques face significant challenges when applied to large 2D geometries. Simulating a 2D system with MPS requires mapping the lattice onto a 1D chain, typically through a snake-like or related ordering. This mapping converts local short-range interactions in the original Hamiltonian into effectively long-range interactions along the 1D representation, substantially increasing the computational cost—particularly for wide or highly connected 2D systems. Moreover, the specific wrapping pattern used in the mapping plays a nontrivial role in determining both the efficiency and accuracy of the simulation. For arbitrary 2D geometries, identifying an optimal ordering that minimizes entanglement growth across MPS bonds is itself a nontrivial optimization problem. As a result, for highly entangled and large 2D circuits, the bond dimension required to achieve accurate results often grows to computationally prohibitive values. An often-used mitigation strategy is to impose cylindrical boundary conditions, treating the system as extended in one direction while keeping a finite circumference in the other. In this setup, the entanglement across an MPS bond scales primarily with the cylinder width rather than the total 2D area, allowing simulations of much longer systems than would be feasible with fully periodic boundary conditions. For this reason, cylindrical geometries are widely employed as a practical compromise between genuine 2D physics and computational tractability. Nevertheless, the required bond dimension still grows exponentially with the cylinder circumference, $\chi \sim e^{S}$, and for highly entangling dynamics or deep random circuits even moderate widths quickly become computationally prohibitive. Thus, while cylindrical MPS can extend the accessible regime, they do not fundamentally overcome the entanglement bottleneck relevant to large-scale quantum-advantage experiments. In the IBM Kicked Ising experiment, the authors estimated that the bond dimension required to exactly represent the stabilizer state and its time evolution in the strongly entangling regime $\theta_h = \pi/2$ at circuit depth 20 would be on the order of $7.2 \times 10^{16}$. They further estimated that storing an MPS with bond dimension $\chi = 1 \times 10^{8}$ would require approximately 400 PB of memory. Similarly, in the D-Wave experiment, the authors projected that MPS simulations would require millions of years of runtime on the Frontier supercomputer, together with approximately 700 PB of storage, to match the quality of the largest QPU simulations. These estimates form a central component of their respective quantum-advantage claims.

For large 2d systems, direct 2D tensor network ansätze, most notably \emph{PEPS}, provide a natural alternative. Unlike MPS, PEPS are intrinsically suited to 2D geometries and do not require mapping the lattice onto a 1D chain. However, PEPS suffer from a fundamental computational bottleneck: the contraction of the tensor network required for evaluating observables cannot, in general, be performed exactly and efficiently, and is known to be a computationally hard problem~\cite{PEPShardSchuch2007, PEPShardHaferkamp2020}. Consequently, one must rely on approximate contraction schemes, such as corner transfer matrix renormalization group~\cite{ctmrg_Nishino1996, ctmrgOrus2009} or boundary MPS methods~\cite{VerstraeteTNreview2008,orus_TNreivew_2014}. Belief-propagation–based approaches adopt the most aggressive simplification, selecting a gauge in which the environment of a tensor (or block) is approximated by a rank-1 object, effectively a product state. More sophisticated variants relax this approximation by representing the environment as a low-rank MPS. Another line of development—exemplified by split-CTMRG techniques~\cite{splitCTMRGXiang2017}—retains separate PEPS layers and reorganizes environment tensors to substantially reduce memory and runtime requirements while preserving high accuracy. These innovations are not merely aesthetic refinements; they constitute essential strategies for rendering large-bond-dimension PEPS contractions computationally feasible. State-of-the-art finite fermionic PEPS simulations with bond dimension $D=28$ have successfully captured the $16\times16$ Hubbard model at 1/8 doping, outperforming leading DMRG approaches on wide ladders~\cite{PEPSD28Chan}. This represents a substantial advance in the expressive capacity of 2D tensor network ansätze.

The \emph{isometric tensor network ansatz}~\cite{ZalatelisoTN2020} was introduced as an attempt to alleviate the contraction bottleneck inherent in PEPS. By imposing local isometric constraints, one can simplify parts of the contraction and improve numerical stability. However, this construction also introduces important limitations. First, the expressive power of isoTN is fundamentally constrained. Any isoTN can be reduced to an MPS by collapsing the network along a row or column direction. As a consequence, two-point correlation functions evaluated along its orthogonality hypersurface must decay exponentially, reflecting the effective 1D character of the ansatz. In contrast, generic 2D tensor network states can, in principle, represent power-law correlations and more complex entanglement structures. Thus, isoTN can faithfully represent only those states compatible with an effective 1D area law, and the imposed isometric constraints further restrict its variational flexibility. Second, maintaining the isometric structure requires dynamically shifting the orthogonality center across the lattice, typically implemented via so-called Moses moves. These transformations introduce additional approximation errors beyond those already present in the underlying optimization schemes (e.g., DMRG- or TEBD-based updates). In the context of the IBM experiment, further limitations arise from the lattice embedding. The authors employed a square-lattice isoTN to simulate the heavy-hexagon architecture, effectively embedding 127 physical qubits into a 195-qubit square lattice by introducing auxiliary (“fake”) qubits. This embedding increases the effective coordination number from the heavy-hexagon average of approximately 2.5 to 4, thereby artificially increasing the simulation cost. The mismatch between the physical hardware connectivity and the ansatz geometry further likely contributed to the inability of isoTN to efficiently and accurately simulate the IBM experiment in the strongly entangling regime. 

Recent developments on \emph{Belief Propagation} have achieved notable success in challenging some of the quantum-advantage claims. Its success can be attributed to two key factors. First, the choice of a tensor network geometry that faithfully reflects the connectivity of the quantum processor plays a crucial role. Second, for the heavy-hexagon lattice and the specific model considered in IBM's Kicked Ising experiment~\cite{IBMkickedisingNat2023}, the correlations in the parameter regimes studied exhibit an underlying tree-like structure. The BP method is particularly efficient and accurate in such settings; in fact, BP becomes exact when the correlations are strictly tree-like. The \emph{graph PEPS} method employed in \cite{PatragPEPS2024}, which uses mean-field environments for both tensor updates and the computation of observables, is closely related to BP. A key distinction is that graph PEPS does not require additional tensor gauging, which can lead to improved computational efficiency. Nevertheless, both approaches rely on 
representing the environment of a tensor as effectively separable—approximated by a product structure. In this sense, both \emph{BP} and \emph{graph PEPS} are uncontrolled approximations, as their validity depends on the absence of strong, nonlocal correlations in the surrounding network. Consequently, these methods are expected to struggle in highly entangled systems whose correlations extend beyond tree-like structures, particularly in loopy lattices with large coordination numbers. Frustrated systems defined on geometries such as triangular or kagome lattices are especially challenging, as geometric frustration and loop-induced correlations undermine the product-environment assumption. Even in simpler lattices, such as cubic geometries with moderate coordination numbers, BP and related methods are expected to deteriorate near quantum critical points, where correlation lengths diverge and entanglement becomes long ranged—behavior that challenges tensor network approximations more generally. It is precisely in such regimes—large 2D systems with strong loop-induced entanglement, high coordination numbers, frustration, and proximity to quantum criticality—that quantum processors may have the greatest potential to outperform classical tensor network methods. By carefully selecting lattice geometries and tuning Hamiltonian parameters to enhance entanglement and long-range correlations, one can push classical approaches toward requiring prohibitively large bond dimensions. In these settings, contraction becomes computationally demanding, and BP-style approximations cease to be reliable, thereby opening a plausible pathway toward demonstrating genuine quantum advantage. At the same time, there has been proposals to improve the BP methods leading to increased accuracy with little to minor simulation costs~\cite{EvenblyLoopBP2024}.

\emph{Operator-based} methods based on Heisenberg time evolution using \emph{MPOs} and \emph{PEPOs} also showed clear advantage in benchmarking the Ising Kicked experiment. Their success can be attributed to the slower growth of operator entanglement entropy observed in \cite{ZalatelMPOIBM2023}, possibly arising from cancellation effects due to the simultaneous application of a unitary gate and its conjugate. The effectiveness of Heisenberg-picture time evolution for this circuit was further demonstrated in \cite{XiangPEPOIBM2023} using projected entangled pair operators (PEPOs). Unlike the MPO approach, the PEPO method does not require an explicit mapping onto a 1D representation of the heavy-hexagon lattice, which constitutes a further advantage. The authors further showed that the PEPO technique can automatically detect low-rank and low-entanglement structures in circuits containing Clifford and near-Clifford gates, thereby reducing computational cost while improving accuracy.

Although several classical techniques have challenged the quantum-advantage claims associated with IBM’s Kicked Ising experiment, it remains an open and intriguing question whether quantum advantage could still be realized on the same—or a closely related—hardware platform by exploring a different dynamical regime or circuit construction. In this spirit, the authors of Ref.~\cite{ZalatelMPOIBM2023} proposed modifications to the original circuit design aimed at increasing classical simulation complexity, for instance by replacing Clifford two-qubit gates with non-Clifford and/or non-commuting interactions. Such changes are expected to enhance operator spreading and entanglement growth, thereby reducing the effectiveness of stabilizer-based or low-entanglement classical methods. However, these proposals naturally raise an important practical question: to what extent can such fine-tuning be implemented within the constraints of existing quantum hardware? Introducing non-Clifford gates or more complex interaction structures may increase sensitivity to noise, calibration errors, and coherence limitations. Thus, while modifying the circuit may render classical simulation more difficult, it may simultaneously impose stricter demands on hardware fidelity and control, potentially narrowing the experimentally accessible regime.

In the D-Wave quantum annealing experiment~\cite{dwavesupremacy2025}, the authors argue that MPS constitutes the only reliable classical tool for benchmarking their quantum processor, effectively dismissing PEPS and NQS approaches. However, it is important to recognize that PEPS are intrinsically tailored for large 2D systems and are, in principle, better suited to such geometries than 1D MPS representations. For relatively small lattices, such as systems up to $8 \times 8$, it is therefore not entirely surprising that MPS performs competitively or even favorably. A more informative comparison would involve identifying regimes in which PEPS begins to outperform MPS and benchmarking those results directly against the QPU data. The authors attribute the shortcomings of PEPS—particularly at longer annealing times—to increasing numerical cost with bond dimension $D$ and to the emergence of system-spanning correlations in slow quenches. However, results reported in Ref.~\cite{Tindalldisordyn2d3d} suggest that extended correlations alone do not fully explain the observed limitations. Instead, the performance of PEPS depends sensitively on the update scheme, gauge choices, and environment approximation. In particular, incorporating belief-propagation–based techniques or improved contraction strategies can significantly enhance both accuracy and efficiency. More broadly, PEPS is widely regarded as the natural tensor network ansatz for 2D and higher-dimensional systems. By relying primarily on 1D MPS methods to benchmark two- and three-dimensional problems, the authors in Ref.~\cite{dwavesupremacy2025} encountered substantial computational bottlenecks. When these classical resource estimates were contrasted with the capabilities of the quantum annealer, they formed the basis of the reported quantum-advantage claims. While MPS can still handle systems up to $8 \times 8$, it is not surprising that it becomes impractical for larger-scale two- and three-dimensional simulations—precisely the regime in which more suitable higher-dimensional tensor network approaches should be considered before drawing definitive conclusions about quantum advantage.

In subsequent developments, early results began to challenge the quantum-advantage claims of Ref.~\cite{dwavesupremacy2025}. In particular, Ref.~\cite{Tindalldisordyn2d3d} demonstrated that two- and three-dimensional tensor network methods, when combined with appropriate gauge choices and advanced variants of belief propagation, can accurately reproduce—and in certain regimes even outperform—the D-Wave quantum annealer. These findings significantly raised the classical benchmark and highlighted the importance of employing geometry-aware tensor network ansätze with carefully optimized environment approximations. At the same time, it is important to emphasize that this work did not address all aspects of the original D-Wave study. For instance, it did not explore the most complex lattice geometries, the largest three-dimensional instances, or the longest annealing times considered in Ref.~\cite{dwavesupremacy2025}. Moreover, it did not reconstruct the full quantum state or compute higher-order observables, such as the fourth-order quantities analyzed in the original work~\cite{DwavecommentBP2025}. Consequently, while BP-based tensor network methods have substantially strengthened the classical baseline, it remains an open question whether these or related techniques can consistently surpass the most demanding regimes explored by the quantum annealer. The possibility of quantum advantage in this context therefore remains unsettled.

For Google's RCS advantage claim~\cite{Googlequantumsupremacy2019}, the challenge to it arose from increasingly sophisticated classical simulation methods that reinterpret sampling as repeated contraction of a circuit tensor network. Early work~\cite{Gray2021hyperoptimized} demonstrated that hyper-optimized contraction strategies—combining hypergraph partitioning, slicing, and globally optimized contraction trees—could reduce the estimated classical cost by orders of magnitude, from 10,000 years to months on supercomputers. Subsequent approaches refined these ideas by improving contraction-path optimization, exploiting noise via controlled truncations, and developing high-throughput sampling techniques such as the “big-head” algorithm, which keeps many qubits open to compute batches of amplitudes efficiently. Large-scale implementations on GPU clusters and supercomputers further reduced runtimes from days to hours, albeit sometimes with approximations, correlated samples, or altered circuit definitions. More recent work achieved increasingly realistic simulations at comparable fidelities using advanced sampling (e.g., importance sampling, MCMC) and aggressive slicing strategies, culminating in 2024 results that generated millions of samples in under 100 seconds—faster than Google’s Sycamore processor for similar fidelity~\cite{zhao2024leapfroggingsycamoreharnessing1432}. Overall, these developments do not invalidate the experiment itself but demonstrate that classical tensor network methods, when highly optimized and massively parallelized, can match or even surpass the originally claimed quantum advantage for this specific task, thereby reframing the claim as a moving boundary rather than a definitive separation.

For the recent quantum echoes experiment, one can start by looking at the error sources in the tensor network simulation, the first potential contribution arises from the sampling procedure itself, namely the use of a limited number of projections. The authors employ rejection sampling to select random projections drawn from a uniform distribution. A more detailed quantitative analysis of how the number of sampled projections affects the variance, bias, and overall performance of the simulation would be valuable in assessing the robustness of this approximation. The second source of error stems from the tensor network contraction carried out over this restricted set of projections. As noted in Ref.~\cite{Google_quantum_echo2025}, the contraction order plays a decisive role in determining the efficiency of the simulation, both in terms of runtime and memory usage. To address this, the authors developed a dedicated TNCO, which searches for efficient contraction paths and provides estimates of the expected runtime. When intermediate tensors exceed available memory limits, slicing techniques are applied up to a prescribed width, and the contraction path is refined using local update rules guided by a Metropolis–Hastings algorithm. Further technical details of the contraction strategy are provided in the supplementary material of the cited work. The TNCO framework was subsequently used to simulate OTOCs via their TNMC approach, implemented on GPUs using NVIDIA’s cuQuantum SDK. Performance was benchmarked against Cotengra~\cite{CotengraGray2021}, a widely used library for optimized tensor network contraction. The authors report improvements in runtime when using TNCO, both with and without explicit memory constraints. 

For tasks involving random circuit sampling, finding the best Tensor network contraction order continues to be a central task and a promising research direction. The story is different for circuits that has a fixed structure, low depth or entanglement where traditional approximation technique schemes can be used. For simple network geometries, the contraction path is straightforward to determine. In these cases, however, the main challenge does not lie in path optimization but rather in the large size of the tensors—or equivalently, the large bond dimensions—that are required to accurately represent highly entangled regimes.
This leads to a different type of computational bottleneck, namely the cost of manipulating very large tensors, which in practice translates into substantial memory usage and heavy CPU or GPU demands for large-scale tensor contractions. While this bottleneck can formally be viewed as a contraction problem, it is important to distinguish it from the challenge of finding an efficient contraction path. 

\subsection{Deciding which tensor network method to use}
\begin{figure}
	\begin{center}
		\includegraphics[width=1\textwidth]{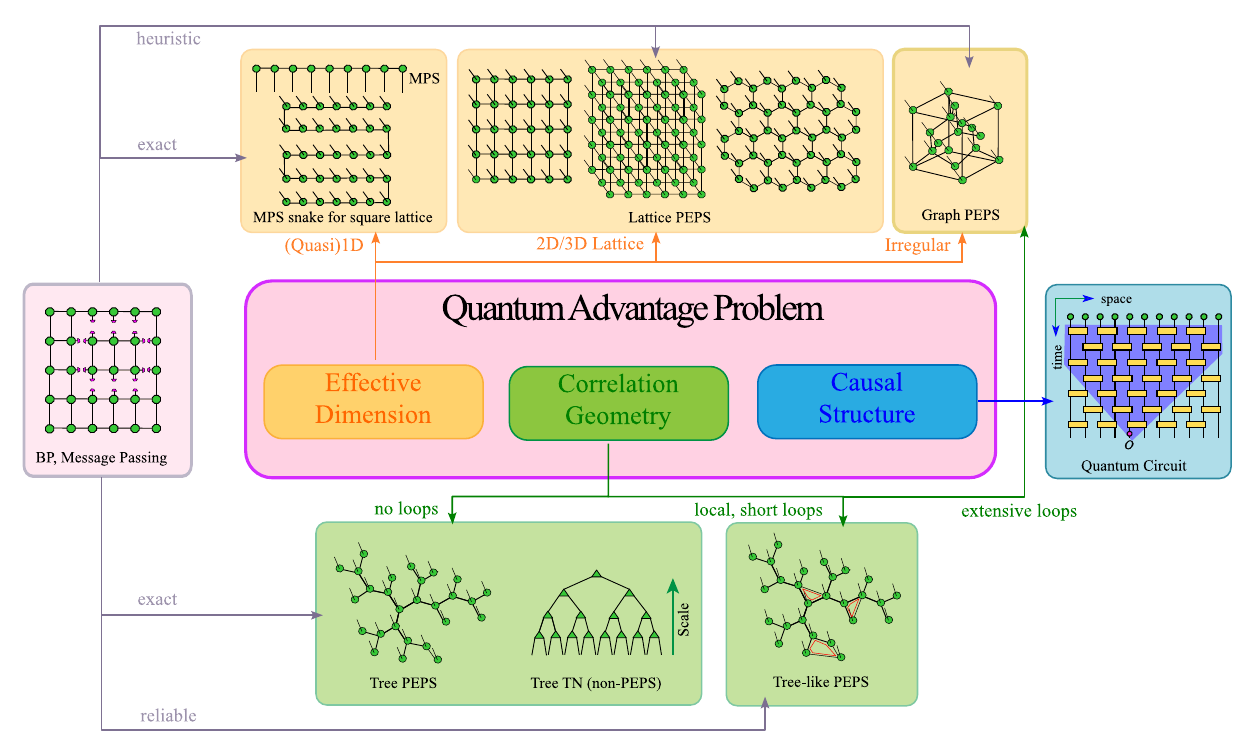}
		\caption{A basic decision tree for selecting an appropriate tensor network method for a quantum advantage experiment. Three main decision axes are illustrated (left): (1) the effective dimensionality of the system (quasi-1D, higher-dimensional lattice, or irregular), which guides the choice of tensor network dimensionality; (2) the correlation structure or geometry (tree, tree-like, or containing extensive loops), which informs the network connectivity and the potential effectiveness of message-passing methods; and (3) the presence or absence of causal structure (eigenstates vs. circuit dynamics), which determines the relevant contraction directions and highlights key physical features such as entanglement growth, area laws, chaos, localization, and integrability.}
    \label{decisiontree}
	\end{center}
\end{figure}

The preceding discussion naturally raises a key and challenging question in the context of quantum advantage experiments: which tensor network framework is best suited for their simulation? To help address this, we present a simple decision tree in Fig.~\ref{decisiontree}, which we hope provides useful guidance, as elaborated below.

Choosing an effective tensor-network method for simulating a quantum-advantage experiment usually requires trying several approaches. Still, a few general considerations are useful. First, MPS-based methods are often viewed as the most scalable and mature. Although they are state of the art for 1D lattice systems, they can remain competitive in higher dimensions. For higher-dimensional problems, PEPS algorithms require careful attention not only to their unfavorable cost scaling but also to errors arising from approximations to the environment tensors. The most accurate and systematic approach is the full-update PEPS algorithm, where the environment of a tensor is approximated using information from (in principle) all tensors in the network. This approach is also the most expensive. In addition, PEPS methods are most developed for regular lattice geometries with small unit cells, since this structure helps control both the cost and the error of environment approximations. For irregular geometries, one can instead use graph-PEPS approaches based on simple updates (with modest extensions such as cluster updates, where the environment includes a fixed-size local cluster). While simple updates scale more easily, they typically introduce poorly controlled errors. Belief propagation (BP) and related message-passing methods can improve the performance of simple-update schemes in this setting.

Incorporating BP also motivates a more refined classification of PEPS by graph structure: (1) tree graphs, (2) tree-like graphs (trees with small, well-separated loops), and (3) graphs with many long loops (regular lattices or generic irregular graphs). The BP approximation is exact on trees, is expected to be reliable on sufficiently tree-like graphs, and is at best a heuristic in the generic case with extensive loops. The best choice of geometry is not always obvious. For clean, translation-invariant systems on regular lattices, it is often natural to define the PEPS on the underlying lattice. For disordered systems or systems on irregular graphs, however, directly mirroring the Hamiltonian geometry may be suboptimal. More broadly, the tensor network should ideally reflect the correlation geometry of the target state, not only the interaction geometry encoded in the Hamiltonian.

\begin{figure}
	\begin{center}
		\includegraphics[width=0.7\textwidth]{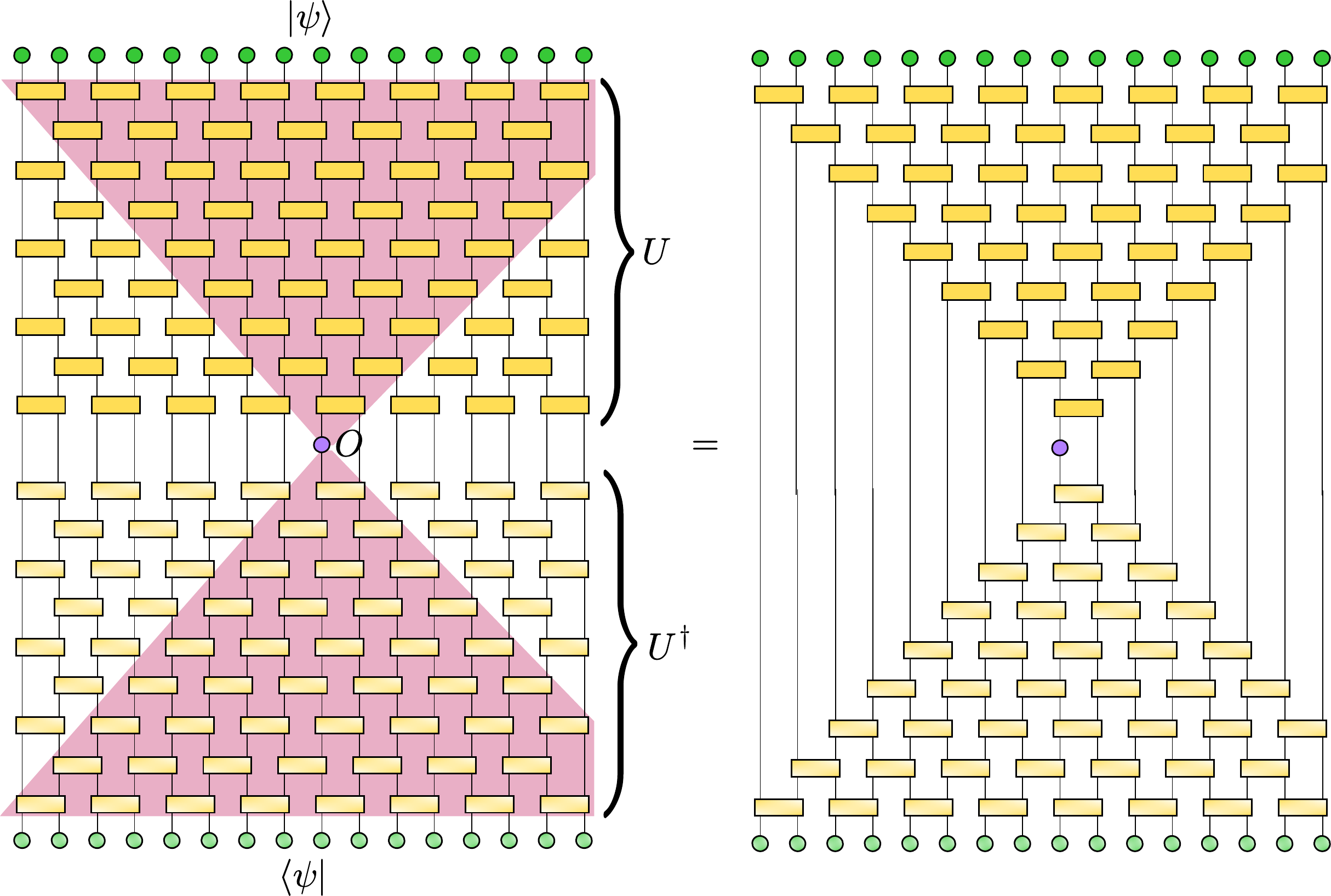}
		\caption{Computing expectation values of local observables in a time-evolved quantum state does not require reconstructing the full final state. Instead, it reduces to contracting a tensor network, where only the tensors inside the causal cone of the operator contribute. The resulting reduced circuit can be contracted in several directions---for example, along the time or spatial directions, or orthogonally to the light-cone boundary, or using coarse-graining methods.}
	\label{fig:lightcone}
    \end{center}
\end{figure}

Another important consideration is whether the target state is static or dynamical. For ground states and low-energy eigenstates of local Hamiltonians, many relevant properties are known—for example, whether the entanglement obeys an area law—and these properties can guide the choice of ansatz. By contrast, time evolution introduces a causal structure. In circuit models, this appears as a light cone that governs how correlations spread and how entanglement grows, see Fig.~\ref{fig:lightcone}. One can exploit this structure by contracting the circuit along different directions (temporal, spatial, or transverse to the light cone), which can yield markedly different intermediate entanglement and computational cost. For dynamical simulations, additional physical features—such as chaos, many-body localization, thermalization, and integrability—can also strongly influence classical simulability.

\subsection{Final discussion: Quantum Computers vs Tensor Networks}
\label{sec:final_discussion}
The ongoing tug-of-war between quantum processors and tensor network algorithms has proven unexpectedly productive. Each advance in a quantum advantage experiment has compelled classical tensor network methods to evolve. Conversely, each time tensor networks narrow the performance gap, they raise the threshold for what constitutes meaningful “advantage.” This iterative dynamic has sharpened both fields, transforming quantum-advantage claims into increasingly rigorous tests of classical and quantum capabilities alike.

Recent supremacy experiments have directly catalyzed major advances in tensor network algorithms. Two developments are particularly noteworthy: the emergence of highly optimized large-scale contraction strategies and the growing integration of tensor network techniques with belief-propagation–based methods. These innovations have enabled tensor network simulations to scale to large and ultra-large high-performance computing (HPC) platforms—an acceleration that was likely driven, at least in part, by the competitive pressure exerted by quantum hardware.

The influence, however, runs in both directions. Tensor networks are now the default classical benchmark against which any new advantage claim must be evaluated. This reality introduces a practical and conceptual challenge: how should an experimentalist identify the appropriate tensor network baseline? The current tensor network landscape is vast and heterogeneous, encompassing MPS, PEPS, tree tensor networks, MERA, hypertree constructions, hypergraph-based contraction optimizers, and a growing array of specialized approximation schemes. The boundaries between these approaches are often blurred, and their effectiveness depends not only on formal complexity considerations but also on implementation details, heuristic optimizations, and architectural compatibility with modern HPC infrastructures such as GPU clusters or TPUs. As a result, selecting a genuinely competitive tensor network challenger has become a nuanced and technically demanding task. The classical baseline is no longer a fixed reference point but a moving target—one shaped by continual algorithmic innovation and engineering refinement.

Taken together, these observations suggest a cautious but important conclusion. In the regime of 200–300 qubits—the likely next frontier for quantum-supremacy experiments—standard entanglement-based classifications offer limited predictive power regarding classical simulability. Tensor network methods are already operating near their practical limits in the 100–200-qubit range for highly entangling circuits. Incremental increases in system size, connectivity, or entanglement structure may be sufficient to shift the balance decisively toward quantum hardware.

In the end, for experimentalists designing future quantum-supremacy demonstrations, the relevant question is not merely which ansatz is formally capable of representing a state, but which method can be optimized, contracted, and stabilized at sufficiently large bond dimension within realistic computational budgets. Classical benchmarking remains indispensable, but its relevance now lies in understanding concrete limits of optimization, contraction cost, geometry, and bond-dimension growth—rather than relying solely on (asymptotic) entanglement arguments.

From the perspective of tensor network methods, achieving a convincing quantum advantage requires identifying regimes in which all leading classical tensor network techniques fail to provide an efficient and accurate simulation. A meaningful discussion of quantum advantage must therefore be grounded in a precise understanding of the structural limitations of these methods. Dynamical scenarios such as quantum quenches, where entanglement can increase rapidly and extensively, remain among the most promising testbeds for probing the limits of classical simulation. Experiments such as IBM’s Kicked Ising study exemplify this strategy, as rapidly growing entanglement can quickly overwhelm fixed-bond-dimension approximations. However, simply targeting regimes of large entanglement is not sufficient. Designing a compelling quantum-advantage experiment requires careful calibration based on a detailed understanding of the specific tensor network methods likely to be used as classical baselines, as well as the architectural constraints of the quantum hardware platform. Critical factors include the choice of Hamiltonian, the parameter regime, lattice geometry, circuit structure, connectivity, and system size. Looking ahead, both classical and quantum investigations are likely to move toward more sophisticated lattice geometries, higher-dimensional systems, and models incorporating longer-range interactions. On the classical side, tensor network techniques will continue to evolve, including improved message-passing schemes, hybrid contraction strategies, and large-scale GPU implementations. The dynamic interplay between algorithmic innovation and hardware advances will ultimately determine the boundary between classical tractability and genuine quantum advantage.

\bibliographystyle{ieeetr}
\bibliography{main}
\end{document}